\documentclass[12pt,preprint]{aastex}
\paperid{58977}
\shorttitle{74~MHz VLA}
\shortauthors{Kassim et al.}

\newcommand{\mjybm}{\mbox{mJy~beam${}^{-1}$}}

\newcommand{\task}[1]{\texttt{#1}}
\newcommand{\aips}{\textsc{aips}}
\renewcommand\anchor[2]{{#2}\footnote{#1}}%
\renewcommand\url[1]{\mbox{#1}}%

\begin{document}
\title{The 74~MHz System on the Very Large Array}
\author{N. E. Kassim, T.~Joseph~W.~Lazio}
\affil{Naval Research Laboratory, Remote Sensing Division, Code~7213,
	Washington, DC 20375-5351}
\email{Namir.Kassim@nrl.navy.mil}
\email{Joseph.Lazio@nrl.navy.mil}

\author{W. C. Erickson}
\affil{U.~of Tasmania, School of Math.\ \& Physics, G.P.O.~Box 252-21,
	Hobart, Tasmania 7001, Australia}

\author{R. A. Perley}
\affil{National Radio Astronomy Observatory, P.O.~Box O, Socorro, NM 
	87801}
\email{rperley@nrao.edu}

\author{W.~D.~Cotton}
\affil{National Radio Astronomy Observatory, 520 Edgemont Road,
	Charlottesville, VA  22903}
\email{bcotton@nrao.edu}

\author{E.~W.~Greisen}
\affil{National Radio Astronomy Observatory, P.O.~Box O, Socorro, NM 
	87801}
\email{egreisen@nrao.edu}

\author{A.~S.~Cohen, B.~Hicks}
\affil{Naval Research Laboratory, Remote Sensing Division, Code~7213,
	Washington, DC 20375-5351}
\email{Aaron.Cohen@nrl.navy.mil}
\email{Brian.Hicks@nrl.navy.mil}

\author{H.~R.~Schmitt}
\affil{Naval Research Laboratory, Remote Sensing Division, Code~7215,
       Washington, DC 20375-5351\\
       and\\
       Interferometrics, Inc., 13454 Sunrise Valley Drive, Suite 240, Herndon,
VA\,20171}
\email{Henrique.Schmitt@nrl.navy.mil}

\and

\author{D.~Katz}
\affil{U.S. Naval Academy, Physics 9C, Annapolis, MD21402-5026}
\email{dkatz@usna.edu}

\begin{abstract}
The Naval Research Laboratory and the National Radio Astronomy
Observatory completed implementation of a low frequency
capability on the Very Large Array at~73.8~MHz in 1998.  This frequency
band offers unprecedented sensitivity ($\sim 25$~\mjybm) and
resolution for low-frequency observations.  The longest baselines in
the VLA itself provide 25\arcsec\ resolution; the system has recently
been extended to the nearby Pie Town antenna of the Very Long Baseline
Array, which provides resolutions as high as 12\arcsec.  This paper
reviews the hardware, the calibration and imaging strategies of this
relatively new system.  Ionospheric phase fluctuations pose the major difficulty
in calibrating the array, and they influence the choice of calibration
strategy.  Over restricted fields of view (e.g., when imaging a strong
source) or at times of extremely quiescent ionospheric ``weather''
(when the ionospheric isoplanatic patch size is larger than the field
of view), an \emph{angle-invariant} calibration strategy can be used.
In this approach a single phase correction is devised for each
antenna, typically via self-calibration; this approach is similar to
that used at higher frequencies.  Over larger fields of view or at
times of more normal ionospheric ``weather'' when the ionospheric
isoplanatic patch size is smaller than the field of view, we adopt a
\emph{field-based} strategy in which the phase correction depends upon
location within the field of view.  In practice we have implemented
this second calibration strategy by modeling the ionosphere above the
array using Zernike polynomials.  Images of 3C sources of moderate
strength are provided as examples of routine, angle-invariant
calibration and imaging. Flux density measurements of a sub-sample
of these sources with previously well determined low frequency spectra
indicate that the 74 MHz flux scale at the Very Large Array is
stable to a few percent, and that flux densities tied to the Baars
et al. value of Cygnus A are reliable to at least 5 percent.
We also present an example of a wide-field
image, devoid of bright objects and containing hundreds of weaker
sources, constructed from the field-based calibration.  The paper also
reviews other practical aspects of low frequency observations, in so
far as they differ from those encountered at higher frequencies,
including aspects of interference excision and wide-field imaging. We
close with a summary of lessons the 74~MHz system offers as a model for
new and developing low-frequency telescopes.
\end{abstract}

\keywords{instrumentation: interferometers --- techniques: high
angular resolution --- techniques: image processing --- techniques:
interferometric --- astrometry}

\section{Introduction}\label{sec:intro}

Radio astronomy began with the discovery of celestial radio emission
by K.~Jansky at~20.5~MHz \citep{j35}.  Throughout the 1950s and 1960s,
key discoveries and technological advances in radio astronomy---at
low-frequencies ($\nu \lesssim 100$~MHz) in particular---helped form
the basis of modern astronomy, including:
\begin{itemize}
\item The introduction of non-thermal processes as an astrophysical
source of emission \citep{ah50,k50,s52}, motivated by early
observations of the diffuse Galactic radio emission at~160~MHz
\citep{r40}, though it took some time for the importance of
non-thermal processes to be accepted widely;

\item The discovery of pulsars through observations at~81~MHz
\citep{hbpsc68}; and

\item The  development of aperture synthesis interferometry at~38~MHz
\citep{rv46,rse50,m52,r52}.  (See also the work by \cite{pp-sm46} and
\cite{mpp-s47} using a ``sea interferometer'' at frequencies
between~75 and~3000~MHz.)
\end{itemize}
By the late 1960s and early 1970s, however, interest turned to
obtaining high-resolution images.  Powerful centimeter-wavelength
interferometers began to provide sub-arcminute angular resolution (or
sub-arcsecond resolution in the case of the VLA by the late 1970s)
with dynamic ranges of several hundred or better.

Even so, a number of low-frequency interferometers continued to be
constructed, some with truly impressive collecting areas ($\gtrsim
10^5$~m${}^{2}$), including the UTR-2 \citep{bms78}, the Culgoora
Radioheliograph,
the Cambridge Synthesis Telescope, and the Clark Lake TPT Radio
Telescope \citep{eme82}.  However, with the exception of the MERLIN
\citep{thom86,leah89} and GMRT 151~MHz
systems \citep{swarup90} and for all telescopes operating below 100 MHz, the
maximum baselines were relatively limited ($\lesssim 5$~km). The corresponding
angular resolution was relatively poor ($\gtrsim 10\arcmin$), and the
resulting high confusion levels meant poor sensitivities ($\gtrsim
1$~Jy).  The primary constraint on baseline length were the phase
distortions imposed by the Earth's ionosphere over the intrinsically
wide fields of view and the lack of suitable algorithms to compensate
for the distortions.  On baselines longer than
a few kilometers, ionospheric phase distortions are severe enough to cause
decorrelation, making higher-resolution imaging difficult to
impossible, especially at the lowest frequencies ($\leq 100$ MHz).

This paucity of large aperture, high-sensitivity, synthesis
instruments operating below~100~MHz has left this portion of the radio
spectrum poorly explored.  Yet, there are a variety of topics that
could be addressed by a sensitive, high-angular-resolution,
low-frequency telescope including:
\begin{itemize}
\item Continuum spectra over much larger frequency dynamic ranges for
studies of shock acceleration and spectral aging in Galactic
(supernova remnants) and extragalactic (radio galaxies and galaxy
cluster relics) sources;

\item Efficient detection of large numbers of steep spectrum sources,
which can be imaged in some cases, including high-redshift radio
galaxies, shocks driven by infalling matter in clusters of galaxies, and
pulsars in the Milky Way and possibly in external galaxies;

\item Probing the ionized interstellar medium (ISM) via measurements
of radio-wave scattering and absorption, the distribution of low-density ionized gas
toward nonthermal sources, and hydrogen and carbon recombination line
observations of very high Rydberg state atoms;

\item The large opacity of \ion{H}{2} regions below~100~MHz can enable
distance determinations to various foreground objects, in both the
Galaxy and external galaxies, from which the three-dimensional
distribution and spectrum of cosmic-ray emissivity can be determined
as well as being used to measure their emission measures,
temperatures, pressures, and ionization states

\item Detection of coherent emission from sources such as the Sun,
Jupiter, pulsars, and possibly radio bursts from nearby stars and
extrasolar planets.
\end{itemize}

While earlier experience with MERLIN and the VLA (and more recently
with the GMRT) demonstrated that large ($>$ 5~km) interferometers could
compensate for ionospheric effects below $\sim$330~MHz, a prototype
74~MHz system was the first to demonstrate that self-calibration
techniques can correct for the large ionospheric-induced phase errors
below 100 MHz. The prototype 74~MHz system consisted of eight of the
VLA's 28 antennas equipped with 74~MHz receivers, and \cite{kped93}
were able to produce images with sub-arcminute resolutions and
sub-Jansky sensitivities, thereby demonstrating that these
self-calibration techniques were able to correct for ionospheric phase
errors on baselines at least as long as the longest VLA baselines
(35~km) and, in principle, much longer.  Because of the limited number
of antennas, the prototype 74~MHz system had relatively poor $u$-$v$
coverage, so only the strongest ($\geq 500$~Jy) sources, numbering a
dozen or so, were imaged.

\begin{figure}
\epsscale{0.7}
{\plotone{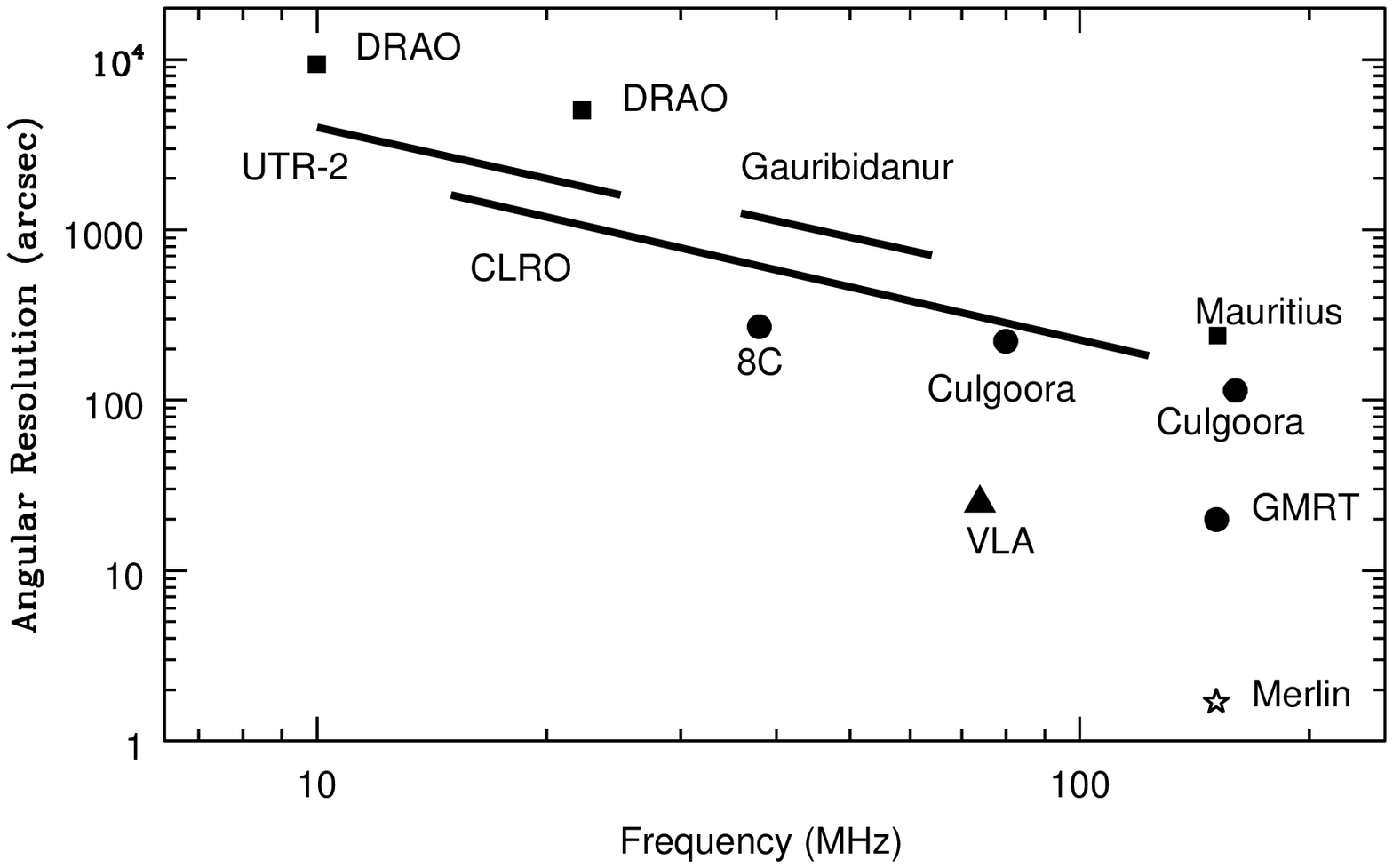}}
\caption[]{Angular resolution (arcseconds) as a
function of frequency (MHz) for past and present imaging
instruments in the 10 to~200~MHz range. The various 
telescopes include the UTR-2 \citep{ks94}, Gauribidanur
\citep{sghs86}, 8C \citep{r90}, Clark Lake Radio Observatory
\citep[CLRO,][]{k88}, Culgoora \citep{sh73,sh75,s77}, Dominion Radio
Astrophysical Observatory \citep[DRAO,][]{bp68,rcs86}, MERLIN
\citep{thom86,leah89} and the Mauritius radio telescope.
The 74~MHz VLA is respresented by the filled triangle.}
\label{fig:resolution}
\end{figure}

\begin{figure}
\epsscale{0.7}
{\plotone{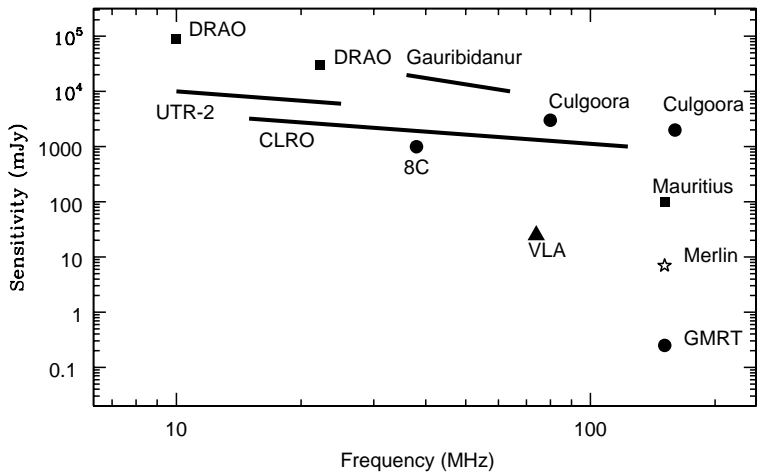}}
\caption[]{Sensitivity (mJy) to a point source as a function
of frequency for the same instruments shown in
Figure~\ref{fig:resolution}.  The sensitivities are estimates of the
minimum detectable flux density provided by past and present
telescopes.  The sensitivity of most of the telescopes shown here were
or are confusion limited; for the VLA and the GMRT, an integration
time of~8~hr was assumed in calculating their sensitivities.}
\label{fig:sensitivity}
\end{figure}

In~1998 January all 28 VLA antennas were equipped with~74~MHz
antennas.  This improved the capability of the instrument greatly, and
it is now possible to detect hundreds of sources in single fields at
high angular resolution and sensitivity.  At the same time, a number
of innovative procedures and software solutions have been developed to
handle the data, some of which are significantly different than at
centimeter wavelengths and others of which are applicable to, and new
to, centimeter wavelengths.  Figures~\ref{fig:resolution}
and~\ref{fig:sensitivity} illustrate the levels of resolution and
sensitivity now possible with this new system, improvements that are
all the more impressive given that the relatively modest collecting
area and low efficiency of the VLA ($\simeq 2 \times 10^3$~m${}^2$
effective collecting area, $\lesssim 10$\% of many of the other
telescopes shown).

This paper describes the fully operational 74~MHz system (hardware and
software) on the \hbox{VLA}.  In \S\ref{sec:history} we summarize
briefly the characteristics of the prototype array as they relate to
the current system.  In \S\ref{sec:general} we review general
characteristics of the current low frequency system, summarize its
general performance, and highlight those aspects that are
significantly different than at centimeter wavelengths.  We describe
the calibration of 74~MHz observations in \S\ref{sec:calibrate} and
their imaging in \S\ref{sec:imaging}.  In \S\ref{sec:dynamic} we suggest
how the current system could be improved via dynamic scheduling.  In
\S\ref{sec:future} we discuss possible future expansions of
low-frequency synthesis instruments using the lessons from the 74~MHz
system, and we present our conclusions in \S\ref{sec:conclude}.
In Appendix~\ref{sec:example} we present selected examples of
imaging of moderately strong 3C sources.

Throughout the paper we shall illustrate various effects with images
or other figures produced from 74~MHz observations.  The examples we
show are a heterogeneous lot, resulting from a number of different
observations of different sources acquired for different purposes.
Our objective is to present a representative sample of various effects,
but not all effects will necessarily be present in every observation.

\section{Low-Frequency Systems on the VLA}\label{sec:history}

The original design of the Very Large Array included only four
frequency bands, centered near wavelengths of~21, 6, 2, and~1.3~cm
\citep{nte83}.  However, there is no fundamental reason a
low-frequency system cannot operate on the array---the principles
of aperture synthesis are as applicable to~50~MHz as they are
to~5~GHz.  More importantly, the key components of the array---the
signal collection (antennas), signal transmission (waveguide), and
signal processing (correlator and post-processing)---are essentially
frequency independent within the radio part of the spectrum.  As soon
as the construction phase of the VLA ended, discussions on
implementing a low-frequency capability began.

\cite{pe84} advocated a free-standing array that would make use of the
VLA's infrastructure (most importantly, the waveguide transmission
system) to achieve approximately 25\arcsec\ resolution.  However, no
source of funding was obvious, and it was decided subsequently that
trial systems could be implemented on the VLA itself to address key
questions regarding the calibration and imaging of low frequency,
long-baseline data.

The initial low-frequency system, operating at~90~cm (300 to~340~MHz),
was installed between~1983 and~1989.  It is a prime-focus system, as
it is impractical to implement a secondary focus system at such a low
frequency.  The feed is a crossed dipole, situated in front of the
Cassegrain subreflector, which thus acts as a (rather imperfect)
ground plane.  Because of this, and because the phase center is
located approximately 50~cm ($\approx \lambda/2$) in front of the true
focus, this system has both a low efficiency (less than 40\%) and a
very broad shoulder of width approximately 12\arcdeg\ in the antenna
power pattern.  Nevertheless, it has been a very successful and widely
used frequency band at the \hbox{VLA}.  Most importantly, it
encouraged the development of the multi-faceted imaging algorithms
\citep{cp92} needed for wide-field, low-frequency observing, as
described later in this paper. Its operation also demonstrated the
robustness of angle-independent self-calibration (\S\ref{sec:pcal}) 
for removing ionospheric distortions across the large ($\sim$2.5 \arcdeg FWHP) 
~90~cm field of view.

The success of the 330~MHz system soon led to consideration of a lower
frequency facility.  A protected radio astronomy frequency allocation
exists between~73 and~74.6~MHz.  Again, funding constraints led to the
decision to deploy a trial system, comprising a simple feed system on
a few of the VLA's antennas.

The feed system chosen is essentially the same as that used
at~330~MHz---crossed dipoles in front of the subreflector.  Because of
the long wavelength, the defocussing errors that affect 330~MHz
performance severely are not serious at~74~MHz.  However, because the
antenna itself is only approximately $6\lambda$ in diameter, the
subreflector is an imperfect ground plane, and the profound effect of
the antenna quadrupod structure, it was anticipated that the forward
gain and sidelobe structure would be fairly poor---as subsequent
measurements have borne out.  \cite{kped93} describe the prototype
74~MHz system in more detail and describe the initial data calibration
and imaging methodologies.

\section{The 74~MHz System on the VLA}\label{sec:general}

Amplifiers and feeds for the complete 74~MHz system were built during
the summer of~1997 by two of us (WCE and BH) at the NRL and deployed
during the fall of that year.  All antennas were equipped with dipoles
by~1998 January.

Because of concerns about blockage at higher frequencies, a deployable
crossed-dipole feed was designed.  The half-wavelength dipoles
contribute to blockage and a higher system temperature, resulting in a
total sensitivity loss of about~6\% at~1.4~GHz and smaller losses at
higher frequencies, so they are deployed only during a fraction of the
time in each configuration.  A simple mounting system is used---two
ropes, each of which supports one dipole, are threaded through
eyebolts located on opposite quadrupod legs at the appropriate height.
The ends of these ropes are tied to cleats located at a convenient
height on the quadrupod legs.  The signal cables drop about~7~m to the
antenna surface, where they pass through the roof of the vertex room
to the amplifiers.  Figure~\ref{fig:mount} shows the dipoles and
mounting system developed.

The receiver units combine the linearly polarized signals of the
dipoles to produce circular polarized signals, then amplify and
bandpass filter these signals, and pass them to the VLA intermediate
frequency (IF) system.  They also contain an integral noise
calibration source.  In order to produce serviceable receivers on a
short timescale and at low cost, they were constructed almost entirely
from commercial components.

The VLA signal transmission system allows for two pairs of two
parallel-hand signals to be transmitted.  The receiver system is
designed so that the two senses of circular polarization from the
74~MHz receivers occupy one pair of signal transmission channels while
the two senses of circular polarization at~330~MHz occupy the other
pair of signal transmission channels.  Thus, 74 and~330~MHz
observations can be acquired simultaneously.  \cite{kped93} used this
simultaneous, dual-frequency capability for ionospheric calibration
via phase transfer from~330 to~74~MHz (see \S\ref{sec:calibrate}).  An
alternate signal transmission approach is to use one pair of signals
for the upper half of the 1.5~MHz bandpass and the other pair for the
lower half, thereby obtaining higher spectral resolution, primarily
for radio frequency interference (RFI) excision purposes
(\S\ref{sec:rfi}). Measurements of circular polarization are normally
available and have been used for both astronomical 
(solar, T. Bastian, private communication, 1998) and RFI excision
purposes.
\footnote{VLA correlator modes 'PA' or 'PB' allow obtaining full 
polarization information at 74 MHz alone, at the expense of halving
the number of channels, and thus reducing the ability to purge 
narrowband RFI. Tests observing a strong, unpolarized source indicate
cross-polarization leakage of at least $\sim$30 \%, and since 
initial attempts at polarization calibration using AIPS task POLCAL
failed, users have not been encouraged to utilize this mode. It is
possible that a dedicated effort with the EVLA correlator might 
permit full polarization astronomical measurements in the future,
although it is suspected that many, except relatively nearby sources,
might be depolarized at this frequency due to Faraday rotation.}

Figure~\ref{fig:receiver} presents a complete block diagram of a
receiver.  In detail, the receiver units are comprised of the
following components:

\begin{enumerate}
\item The two orthogonal linear feeds are converted by a full quad
hybrid 
into right and left circular polarizations (RCP, LCP).

\item An onboard source 
injects a noise calibration signal into both the RCP and LCP chains
via directional couplers.
This source is directly powered by a ``Cal'' signal provided by the site.

\item Out-of-band rejection filtering is provided by high-Q cavity
filters 
with a center frequency of~73.9~MHz and a 1.7~MHz bandwidth.  Such a
narrow bandwidth is necessitated by the close proximity of local
television stations.  (Tests have been conducted with a 3~MHz
bandwidth; the TV signals saturated the receivers.)

\item For the VLA, the signal is transferred directly to the IF
system while for the PT receiver, the signal is upcoverted in order to
pass into the VLBA IF system, by mixing the RCP and LCP channels with
a reference local oscillator (LO) signal.

\item Two power combiners consolidate the 74 and 330~MHz channels for
transport to the correlator.
\end{enumerate}

The initial work on the 74~MHz system focused on the VLA alone.
Concurrently NRAO was in the process of testing fiber optic
transmission techniques by tying in the nearby Pie Town antenna (PT)
of the Very Long Baseline Array (VLBA).  First fringes between a VLA
subarray and the PT antenna were observed in~1998 December, and a
successful test with the PT antenna and the full VLA was conducted
in~1999 September.  Routine observations on this facility began
in~2000 October.  Consequences of this fiber-optic link to Pie Town
(the PT link) are that the longest VLA baselines are extended to
approximately 70~km, but the number of antennas in the VLA decreases
to~26 because the PT antenna replaces one of the antennas in the VLA
and the VLA signal electronics requires removing another antenna from
the array to accommodate the PT signal.

Initial efforts on the PT link focused on frequencies above~1000~MHz.
The VLBA does have a standard observing frequency at~330~MHz, but it
does not have a 74~MHz operating capability.  In~2000, NRL and NRAO
initiated a program to add a 74~MHz capability to the PT antenna and
to operate the PT link at both 74 and~330~MHz.  Initial tests were
conducted with the 74~MHz receiver replacing the 330~MHz receiver, but
the electronics path has been modified subsequently to operate at both
74 and~330~MHz simultaneously.
Figure~\ref{fig:ptlink} shows the first successful fringes at~74~MHz,
obtained on the quasar \objectname[3C]{3C~123} in the fall of~2001.
In the remainder of this paper we shall focus on the VLA system alone.
Many of the techniques we describe are equally applicable to
observations with the PT link, though experience with that system is
considerably more limited and has only been used to observe relatively 
bright, isolated objects \citep{Gizani2005, lazio06, Lane2006, Delaney2004}.
See Appendix~A for an example of the full synthesis VLA+PT image of Cas~A
(Figure~\ref{fig:casapt}).

Table~\ref{tab:perform} summarizes the performance characteristics of
the VLA's 74~MHz system.  We quote a sky-noise dominated system 
temperature that is appropriate for the Galactic polar caps. While
the sky-noise for fields on the Galactic plane can be up to ten times
higher, in practice the low forward gain of the primary beam smooths
out and lowers the variations in T$_{sys}$ to typically a factor
of two or less (\S\ref{sec:ampcal}) The sensitivity
listed is typical for the A and~B configurations, for regions away
from the Galactic plane.  Sensitivities for the smaller configurations
are considerably poorer, because of confusion (\S\ref{sec:confuse})
and presumably because of low-level, broad-band RFI from the antennas
and equipment located at the VLA site.

Figure~\ref{fig:beam} shows the primary beam power pattern measured
for one of the antennas.  Other antennas show similar power patterns.
Table~\ref{tab:perform} cites the primary beamwidth as 11\fdg7, but it
is clear that the beam has only a modest forward gain and a broad
plateau with a poor sidelobe structure.  These result from the
aforementioned aspects of the system that the antenna itself is only
approximately $6\lambda$ in diameter and the antenna quadrupod
structure. Notice that the sharp edge seen on the right side of the
beams presented in Figure~\ref{fig:beam} is due to the fact that this
portion of the beam was not sampled. This was due to motion limitations
of the VLA antennas during the holography measurements used to determine
the beam power pattern. Nevertheless, this sampling issue did not affect
the mapping of the most important part of the beam, down to 20-25~dB
from the peak. The poor primary beam definition gives rise to significant
sidelobe confusion, as discussed below and further in \S\ref{sec:confuse}.

Figures~\ref{fig:bandwidth} and~\ref{fig:time} show the system sensitivity,
as measured by the rms noise level in an image, as a function of increasing
receiver bandwidth and integration time. This behaviour is typical of
observations obtained in the A and B configurations. These images were
excised of narrow-band RFI, calibrated and imaged following the procedures
described in Sections 4 and 5. The deviation from a $\Delta\nu^{-1/2}$
dependence with increasing receiver bandwidth in Figure~\ref{fig:bandwidth}
indicates that the system is not thermal noise limited. On the other hand,
Figure~\ref{fig:time} shows an approximate $t^{-1/2}$ dependence normally
indicative of a thermal noise limited response. Taken together we conclude
that we are 
mainly sidelobe confusion limited since its effects should be independent of
bandwidth to first order, but scale roughly as $t^{-1/2}$ because sources
moving through the sidelobes contribute noise in a random walk fashion that
averages out with time \footnote{One might expect this
dependence to disappear once the $u$-$v$ coverage repeats.  In practice,
however, position shifts caused by the ionosphere (\S\ref{sec:pdelay}) act
to ``thermalize'' the sidelobe confusion contribution.  Thus, we believe
that the noise due to sidelobe confusion will continue to decrease, at least
initially, as $t^{-1/2}$ even after the $u$-$v$ coverage repeats.}
For integrations of $\geq$1~hr, the
deviation from thermal noise is typically a factor of 2-4.

The residual effects of incompletely removed RFI must also play a role in
the reduced sensitivity. In fact the sensitivity in the C and D
configurations is much poorer than expected from confusion alone, and our
hypothesis is that the effects of low-level, broad-band RFI are responsible
for that. In practice the 74 MHz system is rarely used in either the C or D
configurations.

We note that while Table~\ref{tab:perform} quotes an 8 hour sensitivity
limit of ~25 mJy, in practice the achievable sensitivity may vary
considerably due to the positionally dependent sky noise dominated system
temperature, and more importantly from the relative proximity to the handful
of extremely bright sources that often dominate the sidelobe confusion
(e.g. \objectname[]{Cyg~A}, \objectname[]{Cas~A}, etc.)
In \S\ref{sec:confuse} we discuss both sidelobe
and classical confusion further, and show that in the more compact
configurations the latter effect can become significant.

\section{Calibration of 74~MHz VLA Data}\label{sec:calibrate}

Observations of the brightest sources in the sky (e.g.,
\objectname[]{Cyg~A}, \objectname[]{Tau~A}) with the prototype 74~MHz
system demonstrated that the methodologies and algorithms that had
been developed for calibration at the standard VLA frequencies were
generally sufficient for 74~MHz \citep{kped93}, with self-calibration
being particularly important.  The larger number of antennas now
available makes self-calibration even more robust, but it has also
revealed its limits more clearly.

In this section we motivate and describe the procedures we have
developed for the calibration and imaging of~74~MHz data.  The
procedure can be summarized as follows:
\begin{description}
\item[Bandpass calibration (\S\ref{sec:bpass})] Observations at~74~MHz
are acquired in a spectral line mode, both to enable RFI excision and
to avoid bandwidth smearing over the relatively large fields of view.
Therefore we need to apply a baseline correction to the data. 

\item[RFI excision (\S\ref{sec:rfi})] Largely because of
self-interference, 74~MHz data always must be edited to remove
\hbox{RFI}.

\item[Amplitude calibration (\S\ref{sec:ampcal})] As at higher
frequencies, a source whose flux density is presumed to be known must
be observed to set the flux density scale.  The primary flux density
calibrator for the VLA at~74~MHz is \objectname[]{Cyg~A}. Other
sources that can be used when \objectname[]{Cyg~A}
is not available are  \objectname[]{Cas~A}, \objectname[]{Vir~A},
\objectname[]{3C~123} and \objectname[]{Tau~A}.

\item[Phase calibration] The dominant source of phase corruption at
low frequencies is due to the Earth's ionosphere.  Unlike at higher
frequencies, one cannot employ as a phase calibrator a source nearby
in the sky to one's target source or field.  Instead we have developed
two strategies:
\begin{itemize}
\item When the isoplanatic patch (scale over which the rms phase
difference between two lines of sight is approximately 1~radian)
size is larger than the field of view of interest
(\S\ref{sec:pcal}), a single phase calibration can be
applied to the entire field of view.  It is most useful in the more
compact configurations (C and~D) or in the larger configurations (A
and~B) when imaging a strong source.  This strategy relies heavily on
current implementations of self-calibration, and it is similar to the
calibration strategy used by \cite{kped93}. As such, this strategy is
a confirmation of their prediction that self-calibration can be used
to remove ionospheric phase fluctuations.

\item When the isoplanatic patch size is smaller than the field of
view of interest (\S\ref{sec:pdelay}), an angular dependence
\emph{within the field of view} of interest must be used in the phase
calibration. We have used a method called ``field-based'' calibration
to do this, which  models the ionosphere as a phase-delay screen and
uses a grid of background sources to solve for the ionospheric
refraction, both globally and differentially within the field of view.
\end{itemize}
\end{description}
Following sections describe imaging requirements (\S\ref{sec:imaging})
and present examples designed to illustrate the efficacy of both phase
calibration strategies (\S\ref{sec:example}).  This section parallels
closely the discussion in our online
\anchor{http://rsd-www.nrl.navy.mil/7210/7213/LWA/tutorial}{tutorial},
which contains more detailed descriptions of the procedures, as well
as sample inputs for reducing data within \aips.

\subsection{Bandpass Calibration}\label{sec:bpass}

As described below (\S\ref{sec:rfi}), the VLA generates considerable
internal radio frequency interference.  Consequently, observations are
performed in a spectral-line mode, which also avoids bandwidth
smearing over the large regions (typically) imaged
(\S\ref{sec:imaging}).  Characteristic spectral channel bandwidths are
12~kHz in a 1.5~MHz total bandpass.  As with any spectral line
observation, the amplitude variations across the band (and phase gradients
due to delay errors) must be removed by bandpass calibration. Fortunately,
the flux density of \objectname[]{Cyg~A} ($\simeq 17$~kJy) is nearly
always much greater than the equivalent flux density of any RFI, even
in the narrowest channels that the VLA correlator can produce
(12~kHz), meaning that one can use observations of it to calibrate the
bandpass prior to excising \hbox{RFI}.  Other sources---such as
\objectname[]{Cas~A}, \objectname[]{Vir~A}, \objectname[]{Tau~A}---can
also be used; their lower signal-to-noise ratios on longer baselines
can require judicious choices of time or frequency ranges or both in
order to calibrate the bandpass.

Bandpass calibration is traditionally performed with the assumption
that the flux density of the calibrator can be represented by a single
value across the bandpass.  In the case of 74~MHz observations with
the VLA, however, the fractional bandwidth of the system is large
enough that the intrinsic visibility of \objectname[]{Cyg~A} can
change across it.  A rough estimate shows that the effect of
resolution can be ignored if
\begin{equation}
\pi\frac{\Delta\nu}{\nu}\frac{\theta_{\mathrm{src}}}{\theta_{\mathrm{HPBW}}}
 \ll 1,
\label{eqn:bpass}
\end{equation}
where $\Delta\nu/\nu$ is the fractional bandwidth and
$\theta_{\mathrm{src}}/\theta_{\mathrm{HPBW}}$ is the source size
expressed in units of the synthesized beam.  In the A configuration,
the left-hand side of equation~(\ref{eqn:bpass}) is 0.3, so there is a
30\% change (worst case) in visibility across the bandpass on the
longest baseline; in the smaller configurations, the error introduced
by the fractional bandwidth is proportionally less.

Two options exist for dealing with the effects of the intrinsic
visibility change across the bandpass.  First, one can set an upper
limit to the baseline length to be considered, effectively decreasing
$\theta_{\mathrm{src}}/\theta_{\mathrm{HPBW}}$.  Second, one can
divide the observed visibilities at each frequency by a model of the
source, thereby transforming the visibilities into what would have
been obtained had a point source been observed.  In practice, the
second option is used most often in order to employ the maximum amount
of the data possible, and
\anchor{http://rsd-www.nrl.navy.mil/7210/7213/LWA/tutorial/}{models} of
\objectname[]{Cyg~A}, \objectname[]{Cas~A}, \objectname[]{Vir~A},
\objectname[]{3C~123} and \objectname[]{Tau~A} are available online.

This second method, dividing the visibilities by a model of the
source, is done before the visibility amplitudes are calibrated.  At
higher frequencies, the standard bandpass calibration procedure at the
VLA is to divide by the ``Channel~0'' continuum data, formed by
averaging the inner 75\% of the bandpass.  Rather than using a
``Channel~0'' continuum, which would be contaminated by the presence
of the 100~kHz comb (\S\ref{sec:rfi}), one first solves for the individual channel
corrections (both amplitude and phase).  These corrections are then
normalized so that the mean correction across the spectrum is unity.
Various weighting schemes can be used during the normalization
process.  In practice, the weights used in normalization are often
taken to be independent of channel; the converse, scaling the weights
by some function of the amplitude in each channel, can have the effect
of giving more weight to baselines with more RFI (higher amplitudes)
and the short baselines (which are most often affected by RFI).

In general a single bandpass correction is determined from all
observations of the bandpass calibrator, and this correction is
applied to all other observations within that ``observing run.''  For
bright sources, residual bandpass calibration errors can introduce
systematic errors in the image (appearing in the shape of the beam,
but which cannot be \textsc{clean}ed) and thereby limit the dynamic
range of the image.  In these cases, one of two strategies can be
adopted.  If one has multiple observations of the bandpass calibrator,
one can attempt to form the bandpass correction as a function of time.
Alternately, during self-calibration one can produce a different phase
correction for each channel (or small number of channels).

\subsection{RFI Excision}\label{sec:rfi}

Radio frequency interference (RFI) is a significant problem for low
frequency VLA observations, though we note that there is little
\emph{external} RFI within the passband of 73.0--74.6~MHz. RFI is the
major limiting factor for the dynamic range of the data observed with
the more compact VLA configurations (C and D), due to the interference
between the individual antennas of the array. It also constitutes an
important issue for the more extended configurations (A and B),
and needs to be dealt with appropriately.
Essentially all non-astronomical signals are generated internally by
the electronics of the VLA's antennas and (to a lesser degree) signals
emanating from other sources on the VLA site.  By far the most common
is a 100~kHz ``comb'' generated by the VLA's monitor and control
system.  The oscillators responsible for this are located in every
antenna.  The oscillators can be coherent, with low phase rates such
that phase differentials of less than 1~radian are obtained in an
integration time, so
that the coherence is maintained over long periods of time.  The
result is spurious correlation, especially between certain pairs of
antennas whose oscillators appear to maintain coherence regardless of
where they are located in the array.  The net (averaged over frequency)
spurious visibility is not great ($\lesssim 100$~Jy) and can be
effectively removed in the spectral domain.  (By contrast, RFI
at~330~MHz is mainly externally generated but can be excised using
the same procedures as described below.)  Figure~\ref{fig:rfi} (left
panel) shows an example of the 100~kHz ``comb'' for a single baseline.

Various procedures exist to accomplish RFI excision, depending upon
the dynamic range requirements. Notice that the final image dynamic range
will also depend on the position of the source in the field, usually being
lower closer to brighter sources, and how well ionospheric effects can be
solved (Sections 4.4 and 4.5). RFI excision appropriate for moderate
dynamic range images (DR$\sim10^3$) can be accomplished with the following
procedure.\footnote{%
Within \aips, this procedure is implemented using \task{FLGIT} or by
\task{VLAFM}, a special-purpose task added to \aips.
}  One begins by identifying a ``baseline'' region within the spectrum to
which only the astronomical source(s) and the noise within the system
(i.e., $T_{\mathrm{sys}}$) contribute.  Usually, this ``baseline''
region is non-contiguous and does not include the ``comb.''  Next,
visibilities with excessive amplitudes (e.g., $100\sigma$) are
flagged, regardless of where they appear in the spectrum (inside the
baseline region or not).  A linear fit to the ``baseline'' region is
made and subtracted.  The objective of this linear fit is to avoid (or
reduce) the extent to which the flagging is biased by the astronomical
source(s) and the system.  Finally, visibilities whose residual values
exceed a user-defined threshold (typically 6--$8\sigma$) are flagged
in channels both inside and outside of the ``baseline'' region.
The results of this procedure are presented in the right panel
of Figure~\ref{fig:rfi}, where we show the example of the visibility
spectrum on a baseline after RFI editing. An inspection of this image
shows that this procedure has eliminated the RFI emission present in the
left panel.

A similar procedure, also appropriate for moderate dynamic range
imaging, is to flag all of the channels comprising the 100~kHz
comb\footnote{%
Within \aips, the task \task{SPFLG} can be used.
} and then clip any remaining data with excessively large amplitudes.

While elementary, these approaches have proven to be reasonably effective
in removing RFI at~74 (and~330) MHz.  Additional tests also have been
found to be useful.  For instance, channels in which the ratio of the
real part of the visibility to the imaginary part exceeds a given
value can be flagged.  Various observers have noted that RFI can often
be strongly circularly polarized so a test on the amplitude of the
circular polarization (Stokes V parameter) in each channel is also
useful. For the highest dynamic range images (DR$\sim10^4$), it is
often necessary to inspect the data by hand and perform additional
editing.\footnote{ Within \aips, the tasks \task{SPFLG}, \task{TVFLG},
and \task{WIPER} can be used either singly or in combination.} \cite{Lane2005}
describe in more detail post correlation RFI excision with the VLA and VLBA
(see also \cite{Golap2006}).

The limited spectral resolution of the VLA's original correlator means that
even the narrowest channels provided are 12~kHz wide.  As a result, the
procedure we describe can remove a significant fraction of the
data---typically 10 to~25\%.  A key design goal of any future correlator
would be to provide considerably more channels with higher frequency
resolution, allowing much more precise removal of these artificial signals.
Such improvements are forthcoming for the 74 MHz system with the transition
to the EVLA and its WIDAR correlator. Furthermore, techniques of
pre-correlation RFI excision would clearly offer key advances over current
approaches.

A final note is that the discussion above focuses on narrow-band RFI,
much of whose source is relatively well understood. Since the sensitivity
and dynamic range
in the rarely used C and D configurations appears worse than attributable
to confusion or poorly excised narrow-band RFI, we hypothesize that 
an additional form of low-level, broad-band RFI is limiting the system
performance in those configurations. That suggests that 
the source of interference is the VLA itself, or equipment
at the site. Clearly another lesson for future instruments is design them
with a focus on eliminating or at least shielding against self-generated RFI.

\subsection{Amplitude Calibration}\label{sec:ampcal}

\subsubsection{Low-Frequency Calibration Problems and Advantages of the VLA System}\label{sec:ampcal1}

Despite relatively limited sensitivity compared to that typical of
higher frequency observations, the sensitivity of the 74~MHz VLA is
unprecedented for frequencies below~100~MHz, and so it has now become
possible to obtain flux density measurements for thousands of radio
sources that have never been measured before at these frequencies.
However, it is important to note that flux density scales are rather
uncertain below~100~MHz for several reasons. 

First, the Galactic background dominates system noise
temperatures, so system temperatures vary as an antenna
is scanned across the sky and system gain and temperature must be
measured continuously.  Antenna gains vary at less than the few \%
level as a function of time and are consistent with elevation dependent
deformation of the feed or dish structure being negligible at 4-m wavelength.
This trend is consistent with elevation-dependent gain variations at
L-band (20-cm) measured at less than 1\%, while variations at P band
(90-cm) have never been seen. $T_{\mathrm{sys}}$ certainly changes with
time, as it depends on $T_{\mathrm{sky}}$ which is a strong function of
sky position, being much greater on the Galactic plane than off. Early
single antenna tests showed variations of up to a factor of 2 between
``on'' and ``off'' regions against the Galactic plane - being greatest
towards the Galactic center. This is much less than the known contrast
in brightness temperature between inner regions of the Galaxy and cooler
parts of the extragalactic sky as determined by early, lower resolution
measurements (e.g. \citealt{y68}). However that real variation is diluted
by the poor directivity of the 25-m dishes at 74 MHz, with $\leq$20\%
of the power in the main beam and the rest scattered in sidelobes across
the sky. Hence while $T_{\mathrm{sys}}$ does change as a function of time,
the variation is greatly smoothed out and is normally less than a factor
of two over the course of any pointed, full synthesis observations. A
related consideration is that calibrator observations of calibrators like
\objectname[]{Cyg~A} not drive the system response into a regime in which
a non-linear correction
of its correlation coefficient is required. Fortunately the increase on
$T_{\mathrm{sys}}$ when observing \objectname[]{Cyg~A} is modest. A rough
estimate from existing low resolution sky maps suggests an increase in
$T_{\mathrm{sys}}$ of $\leq$40\% in the main beam alone, again consistent
with early single antenna tests. Thus the contribution from the distributed
``hot sky'' still dominates the measured power, and are consistent with the
measurements described below indicating that variations in system gain and
temperature are being tracked accurately (see also Sections 3 and 5.1).

Second, many systems utilize fixed antennas
whose power patterns are not very well determined.  The relative flux
densities of sources at adjacent declinations can be determined
reliably because they traverse the same parts of the fixed antenna's
pattern, but measurements of sources at widely separated declinations
are difficult.  Also, many measurements must be made far from the
maxima of the primary response patterns of the antennas, where the
patterns are less stable than at the maxima.  Finally, ionospheric
amplitude scintillations often disturb measurements below 50~MHz and
ionospheric absorption can affect them below 20~MHz.

The VLA has features that avoid most of these problems.  First, the
gain and noise temperature of each antenna are monitored continuously
by the injection of noise calibration signals into every preamplifier.
At higher frequencies, experience has shown that the gain properties
of the VLA electronics system are stable at the 0.1\% level, and that
linearity is preserved even if the T$_{\mathrm{sys}}$ changes by a
factor of 3 or more.  Second, the VLA antennas are pointed at each
source under observation, and the antenna power pattern is stable,
at least for measurements made near its center.  Sources at different
declinations should be directly comparable.  Third, 74~MHz is a high
enough frequency that ionospheric amplitude scintillations and
absorption are minimal, though occasional  episodes of scintillations
have been seen during observations of strong sources.

\subsubsection{VLA Calibration Method}

Because of the high system temperature of the 74~MHz system, the best
source to use for calibrating the visibility amplitudes is
\objectname[]{Cyg~A}.  However, as it is partially resolved, a good model
is required for accurate amplitude and phase calibration.  We have
used multiple observations in multiple VLA configurations to generate
such a model, and it is available
\anchor{http://rsd-www.nrl.navy.mil/7210/7213/lazio/tutorial/}{online}.
We base our flux density scale on that of \cite{bgp-tw77}, in which
\objectname[]{Cyg~A} has a flux density of~17\,086~Jy at~74~MHz. Although
\objectname[]{Cyg~A} is by far the best amplitude calibrator, other
sources can be employed, if necessary.  \objectname[]{Vir~A} is
acceptable in all VLA configurations, while \objectname[]{Tau~A} and
\objectname[]{Cas~A} can be used in the smaller configurations.  As
with \objectname[]{Cyg~A}, all of these sources are resolved
significantly, so model images (available online) are necessary for
calibration. The amplitude calibration is done in the usual way in AIPS,
with the task CALIB used to derive the antenna complex gains. Residual
amplitude errors are corrected by self calibration in the final stages
of the reduction. As described below, measurements demonstrate that
reliable fluxes, at the 5\% level, can be obtained down to the half power
beam point. In the rare case of severe ionospheric weather conditions
such as ionospheric scintillations (\S\ref{sec:pdelay_motivate}),
amplitude correction is not possible and the data are discarded.

\subsubsection{Reliability of the VLA flux-scale at 74~MHz}

In order to quantify our ability to cope in practice with the problems
described above (\S\ref{sec:ampcal1}), it is necessary to examine the robustness
of the flux density scale for the VLA. As part of a snapshot survey
(\S\ref{sec:example}), we observed a number of strong sources, which
we have used to constrain the gain stability of the 74~MHz system.

\cite{kwp-tn81} presented the spectra of~518 extragalactic radio
sources that have flux densities above~1~Jy at~5~GHz, using
data compiled from many catalogs.  The absolute flux density
scale was based on that of \cite{bgp-tw77}.  Although the data
below~100~MHz are rather sparse, the spectra of \cite{kwp-tn81} have
proven to be quite reliable in most cases.  They do not include any
sources near the Galactic plane and give analytic expressions for
spectra only when they are fairly simple (straight or moderately
curved).  Sources with complex spectra are included in the catalog,
but no attempt was made to fit such spectra with analytic expressions.

Of the 29 sources in our snapshot survey, eleven have spectra for
which \cite{kwp-tn81} give analytic expressions.
Table~\ref{tab:fluxscale} compares our measured 74~MHz flux densities
with the value predicted at~73.8~MHz from those expressions.  (See
also Table~\ref{tab:list}.)

Because of their relatively simple structure even at the highest angular
resolution, the four most reliable sources for comparison appear to be 
\objectname[3C]{3C~98}, \objectname[3C]{3C~123},
\objectname[3C]{3C~219}, and~\objectname[3C]{3C~274}.  These yield an
average flux density ratio (VLA/\citealt*{kwp-tn81}) of~0.99 $\pm$
0.06.  A simple average of all eleven ratios yields 0.98 $\pm$ 0.13.
In either case the agreement between the \cite{kwp-tn81} and VLA flux
density scales appears to be as good as 2\%. We conclude that 74 MHz
flux density measurements are stable at the few percent level, and
that the absolute flux density scale as tied to the \citet{bgp-tw77}
value for Cygnus A is accurate to at least 5\%.  These results also
confirm that antenna system temperatures and gains are being tracked
correctly, and that power is being detected linearly. Hence the 74~MHz
flux scale is reliable for those sources never observed before at this
frequency. However, \cite{kwp-tn81} give correction factors that should
be used to adjust the data from the various catalogs to their flux
density scale. When making comparisons, these corrections should be used.

\subsection{Angle-Independent Phase Calibration}\label{sec:pcal}

The phase calibration procedure employed commonly at centimeter
wavelengths makes use of a secondary calibrator, a modestly strong
source much closer in the sky to the target source than the amplitude
calibrator.  It is chosen to be strong enough to dominate the total
flux within its field of view, so that a unique phase can be derived
for the contribution of the array's electronics.

In contrast, at~74~MHz the field of view is so large that few sources
are strong enough to dominate the field.  Thus, an \emph{initial}
phase calibration is determined from \objectname[]{Cyg~A} (or another
strong source such as those listed above, \S\ref{sec:ampcal}), which
usually provides sufficient coherence on enough short ($\leq 10$~km)
baselines to form an initial image even if the target source is in a
completely different portion of the sky. Oftentimes, coherence is
retained even after the phases are smoothed over long time scales,
even longer than the duration of the observation. 

\emph{This technique works only for bright sources. Implicit in this
procedure is the assumption that a single phase suffices for the entire
field of view---hence its \emph{angle-invariant} designation.}

This initial phase calibration works because relatively larger scales
($\gtrsim 50$~km, i.e., larger than the VLA) present in the ionosphere dominate
the phase fluctuations. It is also the reason previous low frequency
($<100$~MHz) systems with baselines less than about~5~km were able to function
without ionospheric correction techniques---except for a constant refractive
shift arising from the large-scale structures, short baselines ($\lesssim
5$~km) were, to first order, relatively unaffected by the ionosphere
\citep{e84}\footnote{Short baselines are undoubtedly affected by small scale
ionospheric structure beyond simple refractive shifts, but those effects
were negligible for past arrays where the intrinsic sensitivity was already
confusion limited to Jy-levels. As a new generation of low frequency arrays
aims to achieve much greater levels of sensitivity, the phase variations due
to small scale ionospheric structure may well be the limiting factor.}.
The initial image obtained from this calibration must be subsequently
astrometrically corrected and self-calibrated.

We write the observed phase on a given baseline as
\begin{equation}
\phi(t) = \phi_{\mathrm{src}}(t) + \phi_{\mathrm{VLA}} + \phi_{\mathrm{ion}}(t)
\label{eqn:visphase}
\end{equation}
where $\phi_{\mathrm{src}}$ is the phase contributed by the intrinsic
structure of the source, $\phi_{\mathrm{VLA}}$ is the phase
contributed by the VLA electronics, and $\phi_{\mathrm{ion}}$ is the
ionospheric contribution.  We write $\phi(t)$ to emphasize that the
phase is a function of time, not only because the Earth-rotation
synthesis means that the phase can change as the array samples
different portions of the $u$-$v$ plane ($\phi_{\mathrm{src}}$) but 
also because of the temporal variations imposed by the ionosphere
($\phi_{\mathrm{ion}}$).\footnote{%
In practice, $\phi_{\mathrm{VLA}}$ can also be a function of time
because of phase jumps in the electronics.  These are sufficiently
rare that we shall ignore them.%
}  During calibration observations of a strong source, the first
term~$\phi_{\mathrm{src}}$ can be calculated and removed using a model
of the source, so we shall ignore it henceforth.

The smallest scales in the ionosphere over which the rms phase varies
by more than 1~radian at~74~MHz are typically no smaller than 5~km.  As a
result, the ionospheric phase term, $\phi_{\mathrm{ion}}$, can be
considered to be relatively constant across at least the short
baselines in the array ($\lesssim 5$~km).  Transferring the phases
determined toward the calibrator source (i.e., \objectname[]{Cyg~A})
to another direction in the sky introduces a term,
$\phi_{\mathrm{ion,cal}}$, the ionospheric phase distortion toward the
calibrator.  However, for the short baselines, this term is
effectively constant.  It introduces little more than a refractive
shift of the apparent source positions \citep{e84}.  This refractive
shift can be removed by registering the image with an existing all-sky
survey at a higher frequency (i.e., the NVSS, \citealt*{ccgyptb98};
WENSS, \citealt*{rtdmbrb97}; and SUMSS, \citealt{bls99}).

This initial, crude phase calibration is sufficient to produce an
image.  The initial image then serves as the initial model for hybrid
mapping, which consists of iterative loops between self-calibration
and imaging.  With the large number of antennas in the full VLA,
convergence occurs rapidly.  In this respect, the process is similar
to that employed often in very long baseline interferometry (VLBI) in
which a crude initial model is combined with several iterations of
hybrid mapping to produce the final image \citep{w95}.

Phase self-calibration is warranted only if a sufficient
signal-to-noise level can be obtained \citep{cf99}.  In practice there
are two signal-to-noise level thresholds that must be met, but both
can be exceeded quite easily.  The first signal-to-noise threshold
that must be met is that there must be a source or sources that are
strong enough to be detected in the approximate ionospheric coherence
time ($\approx 1$~min.).  Both an extrapolation of higher frequency
source counts \citep{b99} and source counts derived from 74~MHz
observations indicate that there should be roughly 150~Jy of flux
from different sources within the primary beam, originating from
sources stronger than about~5~Jy, which is about~5 times the rms
noise level obtained in a 1~min.\ integration time.

The second signal-to-noise threshold is that the phase derived from
the calibrator source must dominate over weaker sources in the field.
This criterion can be understood by considering the calibrator, with a
flux density~$S_{\mathrm{cal}}$, to be immersed in a ``sea'' of
randomly-located background sources, with a typical flux
density~$S_b$.  Treating the visibilities of the sources as phasors,
the background sources will contribute to a jitter of the phase
determined for the calibrator.  A rough estimate of the magnitude of
this jitter is
\begin{equation}
\delta\phi \sim \sqrt{N}\frac{S_b}{S_{\mathrm{cal}}},
\label{eqn:phasestable}
\end{equation}
where $N$ is the number of background sources in the field of view.
For the purposes of a rough estimate, we take $\delta\phi \sim 0.2$
(implying a phase jitter signal-to-noise threshold of~5).  With $S_b
\sim 5$~Jy and $N \sim 40 (= 200\,\mathrm{Jy}/5\,\mathrm{Jy})$, we
find $S_{\mathrm{cal}} \sim 150$~Jy.  Though clearly a rough estimate,
experience has shown that it is possible to achieve successful phase
self-calibration with as little as 50~Jy from bright sources in the
initial model.  More generally, nearly every randomly-picked field of
view will contain at least one 3C object (or equivalent at southern
declinations) whose flux density alone is close to this minimum value,
and existing all-sky surveys at higher frequencies can be used to
identify the (few) strongest objects totaling at least 100~Jy within
the field of view. (The phase fluctuations in the top panel of
Figure~\ref{fig:ptlink} are consistent with the estimate derived here
for the phase jitter due to background sources.)

A potential weakness of this method is its limited utility to the
larger configurations (A and~B).  Depending upon the state of the
ionosphere, it may be difficult or impossible to obtain sufficient
signal-to-noise on the more limited number of short baselines.  In
turn, this may impair one's ability to produce an initial model for
self-calibration.

An alternate phase calibration strategy does not rely on the phases
transferred from a distant source.  Instead, a strong source, like
\objectname[]{Cyg~A}, is used to establish the amplitude scale.  An
all-sky survey (particularly the NVSS or WENSS) is used to construct a
sky model\footnote{%
Within \aips, the sky model is constructed using \task{FACES}.
} within the primary beam; for sources north of
declination~$+30\arcdeg$, the lower frequency WENSS is preferable to
\hbox{NVSS}.  This sky model then serves as an initial model for
self-calibration.  In effect, no attempt is made to use an external
calibrator to calibrate the phases.  A primary benefit of this
approach is to produce a map with astrometrically correct positions,
as the coordinate system of the map is locked to the \textit{a priori} 
known position of sources in the sky, while self-calibration with a
model produced from raw data locks the final position to an arbitrary
sky position determined by the position of the model or ionospheric
refraction. This alternate strategy
is a simplified application of the field based phase calibration technique,
described in the next section.

\subsection{Field-Based Phase Calibration}\label{sec:pdelay}

The prior calibration strategy is most useful at~74~MHz within a
restricted field of view containing a strong source or strong sources
located relatively close together.  In this section we first motivate
why phase angle independent self-calibration is insufficient for all
observations, then describe the method we have developed to handle
more general observations.

\subsubsection{Motivation}\label{sec:pdelay_motivate}

The isoplanatic patch is the characteristic scale over which the rms
phase difference between two lines of sight is approximately 1~radian.
At low frequencies the size of the isoplanatic patch is determined
primarily by the ionosphere.  The phase contributed by a cold plasma
is
\begin{eqnarray}
\phi &=& r_e\lambda\int_0^D ds\,n_e(s),\nonumber\\
     &=& r_e\lambda N_e,
\label{eqn:plasma}
\end{eqnarray}
where $r_e$ is the classical electron radius, $\lambda$ is the
free-space wavelength, $n_e$ is the electron number density, and $N_e$
is the electron column density (also known as the total electron
column, TEC).  For reference, two lines of sight for which $\Delta\phi
\sim 1$~rad at~74~MHz would differ in electron column density (TEC) by $\Delta
N_e \sim 10^{14}$~m${}^{-2}$.

Two lines of sight separated by the diameter of the primary beam
pierce the F-layer of the ionosphere (the most dense region of the ionisphere
at an altitude of $\sim$400~km) at a linear separation as large
as 80~km.  This is larger than the size of the array itself, even with
the PT link, and the ionosphere can have significant column density
variations at much smaller scales ($\lesssim 10$~km), so the primary
beam typically includes multiple isoplanatic patches.
Figure~\ref{fig:isoplane}, which itself is a portion of a larger
image, shows the effect of constructing an image larger than the
isoplanatic patch.  Clearly evident is a systematic distortion of the
sources attributable to incorrect phase calibration.

Figures~\ref{fig:ionshift}--\ref{fig:ionsmall} illustrate the
successively higher order ionospheric effects that lead to the
break-down of a simple angle-invariant self-calibration.
Figures~\ref{fig:ionshift} and~\ref{fig:ionwedge} shows the effects of
the largest scale ($> 1000$~km) ionospheric structure, a ``wedge''
that acts to shift the entire field of view, without source
distortion, on time scales of minutes. This structure dominates the
total electron content (TEC) and also causes Faraday rotation of
linear polarization. A GPS-based method to correct this effect has
been demonstrated at the VLA \citep{epfk01}, but self-calibration
alone can compensate for it if the time scale on which the phase
corrections are calculated is sufficiently short to track this gross
refraction.  It will, however, leave the field with a gross
astrometric offset from the correct source positions.

Figure~\ref{fig:iontid} shows the phase effects imposed by ionospheric
mesoscale structure.  These mesoscale structures are due typically to
traveling ionospheric disturbances (TIDs), with scale sizes on the order of
hundreds of kilometers, column densities of order $10^{15.5}$~m${}^{-2}$
and periods less than 90 minutes.  While the size of the VLA---even in
its A configuration---is smaller than that of a TID, it is still
sufficiently large that TIDs can impose a (mainly) linear phase
gradient down the arms of the \hbox{VLA}.  The effect of this linear
phase gradient can be seen as a (nearly) linear increase in the phase
with increasing distance from the array center; the amplitude of the
offset also increases with baseline, as the antenna separation becomes
a larger fraction of the \hbox{TID}.  (In the example shown in
Figure~\ref{fig:iontid}, the traveling nature of the disturbance is
also present as a lag between the times when the maximum phase offset
is obtained at the various antennas.)  Self-calibration can remove the
effects of TIDs only for a limited region in the field of view.  At
any given instant, sources outside this region are ``differentially
refracted'' or shifted by varying amounts in proportion to their
distance from nominal direction of the self-calibration solution.
Because a typical observation lasts much longer than the time it takes
a TID to pass over the array, the differential refraction changes with
time.  Thus, sources in a map made using all of the data {are} blurred
as well.

\cite{je92a,je92b} conducted an extensive study of acoustic gravity wave
generated TIDs having phase speeds less than 200~m~s${}^{-1}$ and
periods greater than $10^3$~s using the VLA at~330~MHz.  They found
them to have a quasi-isotropic distribution in azimuth.  On shorter
time scales ($< 300$~s) they found the ionospheric phase effects to be
dominated by faster ($> 200$~m~s${}^{-1}$) magnetic eastward-directed
disturbances (MEDs).  Over the baseline relevant to the extended VLA
configurations and the PT link ($\sim 50$~km), both TIDs and MEDs
contribute to~74~MHz phase effects.

Figure~\ref{fig:ionsmall}a-f were obtained from an 8 hour observation
towards a strong source (Virgo A), and illustrate a variety of effects
due to ionospheric phenomena typically observed at the VLA on scales
sizes both larger and smaller than TIDs. Figures~\ref{fig:ionsmall}~a
and b illustrate first order effects on the visibility phase
towards Virgo A. Figures~\ref{fig:ionsmall}~c and d track the flux density and 
apparent position of Virgo A, while Figures~\ref{fig:ionsmall}~e and f
reflect differential refraction towards five field objects after the
first order term has been
removed. These differential phase effects due to smaller scale ionospheric
structures, on scales of tens of kilometers, are the most intractable and
pose A MAJOR challenge for future instruments. (The high frequency
``jitter'' superposed on the TID-produced structure in
Figure~\ref{fig:iontid} is also contribution from smaller scale structure.)
These phase effects are generated by turbulent structures comparable to OR
SMALLER than the size of the VLA, and lead to source distortion.
There is no current
means of correcting for these effects over a wide field of view and when
they are severe, the data must be discarded.  If these small-scale 
structures generate phase distortions larger than 1~radian on scales
smaller than the size of the VLA antennas, the result is ionospheric
scintillations (also illustrated in Figure 14).  We believe that there
will never be a means for compensating for this effect, at least so far
as imaging applications are concerned, as the data are effectively smeared
in phase before they arrive at the antennas.  (Ionospheric scintillations
can be useful for studying the ionosphere itself, though.) Fortunately 
ionospheric scintillations are rare at the VLA at~74~MHz, but when they
do occur those portions of the data must be removed.

In principle, a data-adaptive calibration scheme, based on
self-calibration, could be used to remove the phase errors that result
from imaging a region larger than the isoplanatic patch.  Current
implementations of self-calibration\footnote{%
This description includes \texttt{CALIB} within \aips.%
} model the phase error on the $i^{\mathrm{th}}$ antenna at time~$t$
as a single quantity, $\delta\phi_i(t)$, equivalent to assuming that
the isoplanatic patch has an infinite extent above the array.  The key
assumption for self-calibration---that errors can be modeled as being
antenna-dependent---remains true for a data-adaptive scheme.  Thus, a
joint multi-source self-calibration, in which the calibration
correction would become a function of position on the sky,
$\delta\phi(t; \alpha, \delta)$, seems possible.  Limited testing
suggests that such a scheme could work, though it probably requires
considerably better signal-to-noise than can be achieved with the VLA;
no comprehensive attempt to implement and assess such a scheme has
been performed yet. In principle such a technique might allow 
wide-field imaging across the full extent of the primary beam
on the longest baselines allowed by natural limits of brightness temperature
sensitivity (for a related discussion, see also \cite{Erickson2006}).

When it is desired to image an entire field of view, the assumption of
an infinite isoplanatic patch is no longer valid. The effects of such
an assumption can be seen as a tendency to detect a larger number of
sources toward the direction of the strongest source in the field
\citep{cohenetal03}. Figure~\ref{fig:SCbias} illustrates this effect
in a field containing \objectname[3C]{3C~63} ($\approx 35$~Jy).
The density of sources across the field is clearly non-uniform.
Besides the effects of anisoplanaticity, which shows up at scales of
5-10$^{\circ}$, the non-uniform density can also arise from two other
related causes. First, if sufficiently high-order ionospheric distortions
are present, they contribute to increasing phase errors at increasing
distances from the effective phase center.  Sources at large distances
will be blurred, thereby decreasing their brightnesses. Second,
Figure~\ref{fig:wander} demonstrates that the apparent source positions
also wander over time. Because the amount and direction of the wander can
vary both over the field of view and with time, sources are smeared further
and their brightnesses decrease further.  Only in the neighborhood of a
strong source are the self-calibration solutions able to track the
ionospheric phase distortions accurately.  If one is interested in
imaging only a strong source, the non-uniform distribution of sources
is usually unimportant.  If a field does not contain a strong source,
the self-calibration solutions represent some ``average'' ionospheric
phase distortion across the field.  Again, depending upon the scientific
problem being attacked, using angle-independent self-calibration may be
sufficient.

\subsubsection{Implementation}\label{sec:pdelay_implement}

Compensation for higher order ionospheric phase distortions is
required \citep{cc02,cotton04}, when the entire field of view is of interest
(e.g., in the VLA Low-frequency Sky Survey \cite{cohen07, cohen06})
or when the object of interest is ``close'' to a strong source.

The latter case happens
fairly frequently given that the typical separation of 3C sources on
the sky is approximately 8\arcdeg, comparable to the size of the VLA
field of view.  Moreover, the dynamic range required to detect or
image a weak source close to a strong source may impose much more
stringent constraints on the need to determine the ionospheric phase
than the approximate 1~radian criterion that we gave at the beginning
of the previous section.

In such cases, we model the ionosphere as a phase-delay
screen.\footnote{%
Within \aips, this procedure is implemented using \task{VLAFM}, a
special-purpose task added to \aips.
}  We return to equation~(\ref{eqn:visphase}), assume that
$\phi_{\mathrm{src}}$ is calculable and can be removed, and expand the
ionosphere term as
\begin{equation}
\phi
 = \phi_{\mathrm{VLA}} + \phi_{\mathrm{ion,lo}} 
 + \phi_{\mathrm{ion,hi}}.
\label{eqn:ioncal}
\end{equation}
Here $\phi_{\mathrm{ion,lo}}$ refers to the phase distortion
introduced by low spatial frequency structures in the ionosphere while
$\phi_{\mathrm{ion,hi}}$ refers to high spatial frequency structures.
In order to make a division between ``high'' and ``low'' spatial
frequencies (or ``long'' and ``short'' wavelengths) we make a
``frozen-flow'' approximation in which the ionospheric structures are
assumed not to change internally over the time it takes for them to be
transported across the array.  Under this assumption, spatial and
temporal scales become equivalent.  The smallest structures ($\lesssim
10$~km) are transported at speeds of order 100~m~s${}^{-1}$ on time
scales of order 100~s.  We consider these to be high spatial
frequencies.  Structures of~100~km or larger are transported across in
time scales of order 15~min.; these are low spatial frequencies.

The large-scale structures are larger than even the maximum baselines
of the A configuration.  We therefore decompose the low-frequency
ionosphere term into low-order Zernike polynomials
\begin{equation}
\phi_{\mathrm{ion,lo}} = \sum_{n=1}^2\sum_l A^l_n Z^l_n
\label{eqn:zernike}
\end{equation}
where $Z^l_n$ is the Zernike polynomial, $A^l_n$ is the coefficient of
that polynomial, and the standard conditions apply that $n \ge |l|$
and $n - |l|$ is even.  The $n = 1$ terms account for the large-scale
refractive shift of the field of view while the $n = 2$ terms describe
astigmatism or differential refraction within the field of view.  The
$n = 0$ term is not used because it represents an overall phase
advance or delay (``piston'') to which the interferometer is
insensitive.  We use Zernike polynomials because they represent a
class of polynomials orthogonal on a circle.  Thus, they are useful
for representing distortions across the aperture of the
array.\footnote{%
The phase-delay screen representation and Zernike modeling of the
ionosphere is not strictly correct as the ionosphere is not in the
far-field of the array.  Nonetheless, we view the Zernike
polynomials as a useful first step in modeling the ionosphere.
}
Table~\ref{tab:zernike} summarizes the polynomials used.  The
methodology is quite similar to that used in adaptive optics systems
in optical astronomy.  One important difference between our use of the
Zernike polynomials and that in adaptive optics, however, is that
these polynomials are used to describe wavefront errors in or near the
aperture for the case of adaptive optics systems in optical astronomy
whereas here they represent errors quite far from the aperture plane.

In order to derive the required corrections, snapshot images of
sources in an ``astrometric grid'' are produced, and the offsets
between the apparent and expected locations of sources in the
astrometric grid are determined.  The snapshots must be formed on
short enough timescales so as to track the ionospheric phase
variations, typically 1~min.\ or shorter.  Both the NVSS and WENSS can
be used to produce this astrometric grid as both are constructed at a
high enough frequency and from a sufficiently large number of sources
that the positions of sources within these catalogs is known to an
accuracy much better than the synthesized beam at~74~MHz, even in the
A configuration.  From the source offsets, the coefficients of the
Zernike polynomials can be found in a least-squares minimization.
Figure~\ref{fig:zernike} shows an example of a (subset) of an
astrometric grid and the resulting $\phi_{\mathrm{ion,lo}}$ over the
VLA for a particular observation.

The snapshot images of the astrometric grid sources are useful only if
two conditions are met.  First, the ionosphere must be stable enough
that sources are not defocused seriously but merely shifted from their
expected positions by refraction.  Second, one must be able to
determine $\phi_{\mathrm{VLA}}$ prior to forming the snapshot images.
As in \S\ref{sec:pcal}, the initial phase calibration is determined
from observations of \objectname[]{Cyg~A}, or another primary calibrator.
For a single source at a
well-known position, one need solve for only the $n = 1$ terms
describing an overall refractive shift.  Correcting for this global
refractive shift should yield $\phi_{\mathrm{VLA}}$.  If
$\phi_{\mathrm{ion,hi}}$ is non-negligible (meaning that sources may
be defocused), it will corrupt estimates of $\phi_{\mathrm{VLA}}$.
Because no phase calibration is performed during this strategy (but
see below), any errors in determining $\phi_{\mathrm{VLA}}$ remain in
the data during all subsequent processing.

Figure~\ref{fig:SCunbias} presents the same field as in
Figure~\ref{fig:SCbias}, only this time calibrated using this
field-calibration strategy.  The density of sources across the field
is seen to be far more uniform.  Figure~\ref{fig:sf} quantifies the
improvement that the field-based calibration; it shows the rms jitter
in the apparent separations of pairs of sources of various
separations.  This rms jitter is a fairly direct measure of the
refraction differences induced by the low spatial frequency
ionospheric irregularities.  For separations greater than 2$^\circ$,
the phase screen corrections dramatically reduce the jitter.  Also
important is that the rms jitter shows no trends as distance from the
phase center increases.

One weakness in our current implementation of this strategy is that
the ionosphere is modeled in a piece-wise fashion in time.  No
``smoothness'' constraint is applied to Zernike models from adjoining
snapshots.  Work is ongoing to rectify this weakness.

In order to assess if (or to what extent) $\phi_{\mathrm{ion,hi}}$ is
corrupting our estimate of $\phi_{\mathrm{VLA}}$, we observe
\objectname[]{Cyg~A} multiple times during the course of an observing
run.  By comparing or, more often, averaging the various estimates of
$\phi_{\mathrm{VLA}} + \phi_{\mathrm{ion,hi}}$, we seek to minimize
the contribution of $\phi_{\mathrm{ion,hi}}$ to our estimate of
$\phi_{\mathrm{VLA}}.$\footnote{%
This process is implemented in the task \texttt{SNFLT}, a
custom-designed task designed to work within \aips\ and available from
the authors upon request.
}

We close this section with a few general comments on our choice of
Zernike polynomials and our implementation.  From the work of
\cite{je92a,je92b} showing that much of the ionospheric structure
above the VLA is in the form of waves, one might wonder if a Fourier
representation would not be more appropriate. We have chosen to use
Zernike polynomials to describe the ionospheric structures precisely
because they were invented for the purpose of describing phase errors
across a circular aperture.  Moreover, compared to a rectangular
Fourier transform, many fewer terms of Zernike polynomials are
required to describe the ionospheric phase fluctuations.  Given the
limited sensitivity of the VLA, this criterion is quite important.
Finally, although it is not yet contained within our implementation, a
natural extension of our method would involve requiring the
ionospheric phase corrections to be smooth in time.  In order to
impose this requirement, one requires an orthogonal basis for the
modeling since the interpolation in time is by interpolating the
coefficients and this only works if the terms being interpolated are
orthogonal.

We emphasize that our division between $\phi_{\mathrm{ion,lo}}$
and~$\phi_{\mathrm{ion,hi}}$ is phenomenological and has the effect of
making the division based on the order of the Zernike polynomials used
rather than on physical properties of the ionosphere.  The range of
spatial scales in the ionosphere implies that one could use
higher-order Zernike polynomials to decompose the phase distortions.
In principle, by incorporating a sufficient number of Zernike
polynomials one could reduce $\phi_{\mathrm{ion,hi}}$ to a
sufficiently low level so as to be unimportant; orders as high as 80
are not unusual for correcting large optical telescopes.  However, the
number of Zernike coefficients that can be determined is limited by
both the sensitivity of the system and the \emph{aperture}
distribution of the \hbox{VLA}.  The sensitivity of the VLA is an
issue because the field must be imaged on less than the ionospheric
coherence time, which in turn depends upon the array configuration as
the amount of phase contributed by the ionosphere can depend upon the
baseline length (Figure~\ref{fig:iontid}).  In 1~min., a coherence
time appropriate for the B configuration, a typical field of view
contains no more than 15--20 sources strong enough to be detected; for
A configuration, a more typical coherence time is approximately 20~s.
Even during times of ionospheric quiescence, fewer than 10 sources may
actually be detected; during poor ionospheric ``weather'' conditions
or when sidelobes from strong sources obscure weaker sources, the
actual number detected with any confidence may be no more than 5
sources.  Hence, in order to avoid (or minimize the occurrence of)
spurious Zernike coefficients, we have restricted the modeling to only
the $n < 3$ terms.  This model of the ionosphere is constructed anew
every 1--2~min.

The aperture distribution (as opposed to the more traditional concerns
in interferometry regarding the $u$-$v$ distribution) is related to
the range of spatial scales sampled in the ionosphere.  
In order to characterize large-scale ionospheric structures (e.g.,
TIDs) that are of concern under typical ``weather'' conditions, only a
modest number of pierce points through the ionosphere are required.
In this respect the aperture distribution of the compact
configurations (C and~D) provides adequate sampling of the relevant
spatial scales, though such sampling is also often not needed because
the array remains nearly coherent in these configurations.  In
contrast, in order to characterize smaller scale ionospheric
structures, a high density of closely spaced pierce points is
required, which in turn requires high spatial frequencies in the
aperture plane, rather than in the $u$-$v$ plane.  In the extended
configuration (A and~B), the sparseness of the aperture distribution
means that small spatial scales are hardly sampled at all. The design of
future low frequency instruments may require a compromise between the
good uv coverage important for imaging and the good aperture plane 
coverage that might be required to allow sufficient modelling of small
scale ionospheric structure.

\subsection{Phase Transfer}\label{sec:ptransfer}

Under especially severe ionospheric weather conditions (with
ionospheric phase rates on long baselines in excess
of~1~deg~s${}^{-1}$), it may become necessary to scale the 330~MHz
ionospheric-induced phase rates, transfer, and remove them from the
74~MHz data stream \citep{kped93}.  In practice, this dual-frequency
ionospheric phase transfer technique has been required only rarely, as
even in the A configuration phases transferred from a strong source
anywhere in the sky retain sufficient coherence on enough short
spacings to provide an initial model for self-calibration.  Subsequent
iterations of self-calibration then improve the phases on the longer
baselines. As with straight-forward self-calibration, phase transfer
does not compensate for the main failing of self-calibration, which is
the lack of an angular dependence on the antenna based phase
solutions. However phase transfer may well be of great benefit to future
low frequency instruments, especially those planned to operate at lower
frequencies and longer baselines than the VLA, such as the LWA 
\citep{Kassim2004,Kassim2006} and LOFAR (Kassim et al. 2004). Therefore it is
important that their design does not preclude the possibility of
simultaneous observations at multiple frequencies.

\subsection{Evolving Techniques of Phase Calibration}

The previous sections illustrate successive schemes of ionospheric
phase calibration that continue to evolve with real observational
experience. A technique such as joint, multi-source self-calibration
is required to realize the full potential of emerging larger
low-frequency instruments such as the LWA and LOFAR but has yet to
be developed and tested. Other related appoaches to the calibration
of large, low frequency arrays continue to be proposed and discussed
\citep{Erickson2006,Nijboer2006,Cotton2006,Brentjens2005,Noordam2004,Lane2004}.

Table~\ref{tab:ioncalib} presents a simple overview of these evolving
schemes; the illustrations in Cotton et al. (2004) may help
orient the reader with respect to the applicable geometries.

\section{Imaging at Low Frequencies}\label{sec:imaging}

In this section we summarize the essential procedures needed for
effective imaging in the presence of effects that are particularly
significant at these low frequencies.

\subsection{Confusion and Sidelobes}\label{sec:confuse}

A key limitation to previous low frequency interferometers operating below
100 MHz has been the poor and confusion-limited sensitivity that arose from
their low angular resolution.  The high angular resolution of the larger (A-
and B-) configurations of the VLA provides some mitigation, but the poor
forward gain of the antennas increases the confusion.  In fact, two effects
are at work, and we distinguish between \emph{classical} confusion and
\emph{sidelobe} confusion.  The former occurs when the density of sources
within the synthesized beam becomes so large that they cannot be separated
(e.g., 1 source per every 10~beams is a common criterion for the onset of
classical confusion).  The latter results from the incompletely removed
sidelobe response to bright sources.

Table~\ref{tab:confuse} summarizes the classical confusion limits for
the four VLA configurations, where we have taken classical confusion
to occur when there is one source per 10 synthesized beams.  In order
to estimate these classical confusion limits, we have adopted a $\log
N$-$\log S$ relation derived from existing 74~MHz observations,
\begin{equation}
N(>S)
 = 1.25\,\mathrm{deg}^{-2}\,\left(\frac{S}{1\,\mathrm{Jy}}\right)^{-1.25}
\label{eqn:counts}
\end{equation}
where $N(>S)$ is the areal density of sources, in units of
deg${}^{-2}$, stronger than $S$~Jy.  Strictly, this $\log N$-$\log S$
relation is valid only for flux densities $S \gtrsim 0.25$~Jy, based
on the flux densities of the sources from which it was determined.
Expectations of source densities, based on higher frequency source
counts, suggest that this $\log N$-$\log S$ relation will tend to
overestimate the confusion flux density levels, though, and estimates
using source counts scaled from higher frequencies give similar
results.  Even with the uncertainty in the low-frequency $\log
N$-$\log S$ relation, the results presented in  Table~\ref{tab:confuse}
show that classical confusion is unlikely to limit the sensitivity
of~74~MHz observations in the A configuration and probably not in
the B configuration (possibly only in crowded regions like the Galactic
center), but it is a serious factor for C- and D-configuration observations.

Sidelobe confusion results from the improperly subtracted response to bright
sources both inside and outside the main field of view, and is exacerbated
when those sources are unresolved by the synthesized beam. The dominant
effect is often from the pathlogically brightest sources (e.g. \objectname[]{Cyg~A}, \objectname[]{Cas~A},
etc) outside the main field of view whose emission ``rumble'' through the
primary beam sidelobes.  This problem is particularly severe for the 74~MHz
VLA because of the relatively small aperture ($\approx 6\lambda$) and the
sidelobes caused by the feed support structure. Thus the primary antenna gain is low, the primary beam is large (FWHM
$\approx 11\arcdeg$), and the sidelobe levels are high and asymmetric (peak
sidelobes are typically only 20~dB down from the main lobe), with a large
area of sky in the ``close-in sidelobe'' region (Figure~\ref{fig:beam}).
Moreover, the receiver noise is a small fraction of the total system
temperature, so objects all over the sky, including those seen through the
sidelobes, produce measurable coherence.  Because of this ``signal-rich''
environment at~74~MHz, it is often necessary to image the entire primary
beam and a small number of sources outside it.

Competing factors mitigate the effects of sidelobe confusion, including 
delay beam and "ionospheric" smearing of the residuals of subtracted 
sources. Nevertheless remaining artifacts from the few, strongest 
sources outside the main field of view often limit the sensitivity and 
dynamic range. Sidelobe confusion can still dominate over classical 
confusion even in the more compact configurations, since in those 
cases the arc-minute size scale of the normally offending sources 
(most notably \objectname[]{Cyg~A}, \objectname[]{Cas~A}, 
\objectname[]{Vir~A}, \objectname[]{Her~A}, \objectname[]{Hyd~A}, and 
\objectname[]{Tau~A}) prevents them
from being resolved out, and hence their residual effects are worse. (The
dominance of sidelobe confusion over thermal noise was indicated earlier in
Figures 7 and 8 and discussed In \S\ref{sec:general}. As noted earlier, the
sensitivity in the compact configurations is worse than attributable to
either form of confusion alone, and our hypothesis is that low-level,
broad-band RFI, possibly self-generated, are limiting the performance.

This situation is in direct contrast to that at centimeter wavelengths
\citep[viz.\ Figure~15]{ccgyptb98} where the receiver temperatures
dominate the sky temperature, a more uniform aperture illumination
produces lower sidelobes, and the non-thermal spectra result in most
sources being fainter than at low frequencies, so that only rarely
does one have to contend with sources outside the primary beam.

\subsection{Wide-field Imaging}\label{sec:widefield}

The standard two-dimensional Fourier inversion of visibility data
requires that $w\theta^2 \ll 1$ where $w$ is component of the
interferometric baseline in the direction of the source and $\theta$
is the field of view.  This assumption is not valid for 74~MHz
observations over any significant hour angle.\footnote{%
In general, snapshot observations are not viable with the 74~MHz
\hbox{VLA}.  The large primary beam and crowded fields means that a
snapshot observation produces too few visibilities for an adequate
representation of the field of view.  The exception is for sources
that are so strong, e.g., \objectname[]{Cyg~A}, \objectname[]{Vir~A},
\objectname[]{Cas~A}, that their flux densities dominate the flux
densities of other sources in the field of view.%
}  Consequently, inverting the visibility data requires either a
three-dimensional Fourier inversion or ``polyhedral imaging,'' in
which the field of view is tessellated into facets over which the
assumption of a two-dimensional inversion is valid (e.g.,
Figure~\ref{fig:SCunbias}).  For further details and examples, see
\cite{cp92} and \cite{p99}.  In general, polyhedral imaging is the
more common technique.\footnote{%
The \aips\ task \texttt{IMAGR} and the special-purpose, low-frequency
imaging task \texttt{VLAFM}  implement polyhedral imaging.
}  Table~\ref{tab:facet} gives general guidelines as to the size and
number of facets required for acceptable polyhedral imaging over the
entire primary beam.  In constructing Table~\ref{tab:facet} we have
made use of the criterion \citep{tms86,p99} that the radius of a facet (in
radians) should be
\begin{equation}
\theta_{\mathrm{facet}} \approx \frac{1}{3}\sqrt{\theta}
\label{eqn:facet}
\end{equation}
where $\theta$ is the (FWHM) diameter of the synthesized beam ($\sim
\lambda/b$ for a baseline~$b$).  Larger facets can be used to reduce
the computational burden at the expense of increased phase errors at
the edges of the facets.

An alternate strategy, ``targeted faceting,'' exploits two aspects of
the radio astronomical sky.  First, the majority of the sidelobe
confusion results from the strongest sources either inside or outside the
primary beam.  The variance in the flux density is given by the
expectation value of $S^2 N(S)$, which because $N(S) \propto S^x$
with $x < 2$ (equation~\ref{eqn:counts}) means that the variance
is dominated by the strongest sources. Second, most of the the sky
remains largely dark (empty of sources), even at these frequencies.
Based on these facts, instead of tessellating the entire primary beam,
which would take a large amount of time and computer power, one
makes use of \emph{a priori} knowledge of the field of
view, e.g., from the NVSS or WENSS, to place small facets only where
there are sources. We usually have a few hundred small facets per pointing.
\footnote{ Determining the location of the facets can be done by the \aips\
task \texttt{SETFC} or the special-purpose, low-frequency imaging task
\texttt{VLAFM}.} Depending upon the sensitivity of one's image and the
desired image, this strategy can produce a useful computational savings.
Recently, a new technique for wide-field imaging know as the "w-projection"
has been proposed. It projects the intrinsically 3-D data onto a 2-D plane 
with an appropriate Fernel-like convolution, and avoid the teidum involved
with polyhedron imaging and targetted facetting. This technique is currently
being tested at the VLA \citep{Cornwell2006}.

\subsection{Astrometry}\label{sec:astrometry}

Even with the relatively large synthesized beam, compared to that
attainable at higher frequencies, accurate astrometry at~74~MHz is
desired.  The astronomical motivations are varied but include spectral
index studies, for which one wants to align images obtained at
different frequencies (to the extent allowed by any
frequency-dependent shifts within the source), and followup at other
wavelengths, for which positional accuracies of 1\arcsec\ or better
can be required.

If self-calibration is used during phase calibration
(\S\ref{sec:pcal}), self-calibration will ``freeze'' the large-scale
refraction but leave the image with an arbitrary absolute position (as
in Figure~\ref{fig:ionshift}).  The NVSS or WENSS sources that appear
in the image can then be utilized to re-register the astrometry to an
accuracy of approximately 5\arcsec.  If the ionosphere is treated as a
phase-delay screen (\S\ref{sec:pdelay}), the astrometric source grid
used in the procedure results in source positions comparable to that
of the survey from which the astrometric grid was constructed (e.g., a
few arcseconds for NVSS), provided that there are no other systematic
effects.

\section{Solar Effects, Ionospheric Weather, and Dynamic
	Scheduling}\label{sec:dynamic}

A potential operational improvement in conducting 74~MHz observations 
(and low-frequency observations in general) would be to determine when 
ionospheric conditions are sufficiently quiescent to make useful 
observations.  Currently, observations are scheduled on the telescope 
well in advance and proceed at the scheduled time.  This procedure 
carries the risk that the ionosphere could be in a sufficiently 
disturbed state so as to preclude useful observations.  A better 
approach would be to schedule the telescope in a ``dynamic'' fashion. 
In order to do this, one would have to conduct a test or identify a 
proxy observable that could establish the state of the ionosphere 
rapidly.  If the ionosphere was disturbed so that 74~MHz observations 
would be unlikely to be successful, observations at a higher frequency 
could proceed with the 74~MHz observations being deferred until such 
time as the ionosphere is more amenable to correction via methods 
described here.\footnote{A similar observing strategy is in place for 
high frequency observations on the VLA (22 and 43 GHz) for which the 
dominant source of phase fluctuations is the troposphere.  In this 
case, if the tropospheric conditions appear poor, the high frequency 
observations can be deferred with lower frequency observations being 
observed instead. }  Our experience suggests a number of ways in which 
such dynamic scheduling could be implemented.

In general, the ionospheric phase effects that dominate calibration 
issues at low frequencies are a manifestation of highly unpredictable 
ionospheric weather conditions. The key consideration for low frequency 
observations is not the total thickness of the ionosphere, as measured 
by the TEC, but variations in TEC such as TIDs and smaller scale 
ionospheric structures (\S\ref{sec:pdelay_motivate}).  Stable 
ionospheric conditions can occur at mid-day, when the TEC is highest. 
At the same time, highly disruptive ionospheric scintillations can 
occur during the middle of the night. Periods of particular ionospheric 
instability are [after] sunset and
especially
sunrise (Figure~\ref{fig:ionsmall}), but it is often easy to avoid 
observations during these times.

Solar activity can affect 74 MHz observations both directly, due to 
radio
emission
from the sun that occur during the daytime,
and indirectly, from ionospheric turbulence generated by manifestations 
of space weather linked to solar activity that can occur any time of 
the day or night.

At~74~MHz the quiet \objectname[]{Sun} is a benign
2~kJy disk (compared to \objectname[]{Cyg~A} at~17~kJy), and, at 
30\arcmin\ or larger in size, is significantly resolved out with the A 
and~B configurations.  However, nonthermal solar radio noise [storms] 
can have flux densities in excess of~1~MJy, and by entering through the 
far-out sidelibes of the primary beam render observations completely 
incapable of being calibrated. However they are usually short-lived, on 
the order of 30 minutes or less, and  can be excised from the data
in way similar to the treatment of narrow-band RFI.

The second direct effect is due to the scattering effects of the solar 
wind. These lead to asymetric angular broadening and distortion of the 
brightness distribution of observed sources, and the effects are worst 
in the A and B configuration when the angular resolution is highest. 
Experience with both of these direct effects indicates that a useful 
rule of thumb is to allow at least 60 degree stand-off betweeen the 
target source and the sun.

The indirect effects are more important since bad ionospheric weather 
conditions related to a geomagnetic storm precipitated by solar 
activity can persist for days and render observations useless 
throughout. One well known manifestation are the massive solar 
ejections of plasma and magnetic field known as Coronal Mass Ejections 
(CMEs). Energetic Earth-ward directed CMEs
take
20 to 48 hours to arrive and, given a suitable magnetic field 
orientation,
their impact
can transfer massive amounts of energy into the upper atmosphere. The 
resulting geomagnetic storms can damage satellites, knock out power 
grids, and disrupt communications. Ionospheric disturbances of 
terrestrial origin, for example acoustic gravity waves can also 
generate ionospheric
variations and poor low frequency
observing conditions. However the most severe ionospheric disturbances 
are usually associated with solar activity.

Various spacecraft and ground-based observatories monitor the Sun, and 
indices exist to quantify the level of solar and geomagnetic activity.
Hence proxies exist to
predict both direct and indirect effects of solar and geomagnetic 
activity, and in principle these could be used to provide an 
"ionospheric weather" forecast to guide low frequency observations. 
However no systematic effort has been conducted yet to assess the 
correlation between various indices and satisfactory ionospheric 
imaging conditions. In the mean time, the self-calibration solution, or 
short time scale stability of the phase on a few baselines from a short 
scan on a strong 74~MHz source (\objectname[]{Cyg~A}, 
\objectname[]{Vir~A}) is the most useful means of determining whether 
ionospheric weather conditions are suitable for observations. If they 
are deemed too poor to observe, the observations should be postponed, 
and conditions revisited on the timescale of several hours, or sooner 
if monitoring of strong sources can be efficiently built in to a 
dynamic scheduling system.  Only limited attempts have been made to 
quantify the usefulness of this latter approach; while this procedure 
appears useful, it may not be able to distinguish excellent ionospheric 
imaging conditions from only fair conditions.

\section{The Future of the 74~MHz System}\label{sec:future}

The 74~MHz system expanded from an initial trial system to an
operational component of the \hbox{VLA}.  Even so, its sensitivity is
limited fundamentally by the VLA's modest collecting area ($\sim 10^4$~m${}^2$,
which translates to $\sim2\times10^3$~m$^2$ given the $\sim$15\% aperture
efficiency at this frequency) and poor sensitivity. Even if the VLA
telescopes could be made 50\% efficient, and the bandwidths increased
significantly (\S\ref{sec:general}), it is not possible for the VLA
(or the EVLA or the GMRT) to be more sensitive in plausible integration 
time ($\lesssim$ 10 hrs) than roughly 1~\mjybm\ at 74~MHz---100 times less
sensitive than at~1400~MHz to normal-spectrum objects (i.e., those with
spectral indices $\alpha < -0.7$, $S_\nu \propto \nu^\alpha$).

The only way to achieve a sensitivity at these wavelengths comparable
to that available at centimeter wavelengths is to build a system with
much greater collecting area. The Long Wavelength Array (LWA\footnote{lwa.unm.edu},
\citealt{kassimetal06, Kassim2006, Taylor06, KE98}), Low-Frequency Array
(LOFAR\footnote{www.lofar.org}, \citealt{kassimetal00, Kassim2004, Falcke2005})
and Mileura Widefield Array (MWA\footnote{www.haystack.mit.edu/ast/arrays/mwa/},
\citealt{Morales2006, bowmanetal07}) seek to accomplish this at different
levels, with different collecting areas, spatial resolutions and frequency
ranges. The objective of these projects is to produce
aperture synthesis instruments, operating at frequencies between~10
and~300~MHz, with collecting areas of up to $\sim10^6$~m${}^2$ at~20~MHz with
maximum baselines of up to 500~km or longer. As such, they would have
considerably more collecting area than the VLA (and more than any
other previous low-frequency telescope as well) and angular
resolutions comparable to that of the VLA at frequencies near~1~GHz.

These projects must take into account the lessons learned from previous
low frequency instruments. Here we emphasize some of these lessons, which
became apparent during the deployment of the 74~MHz VLA system:
\begin{itemize}
\item The system should be designed with careful attention to the
signals it emits so that the instrument does not pollute itself with
\hbox{RFI}.

\item The complex gain and noise temperature of each antenna should be
monitored continuously.  Measurements of the round-trip phase through
the system should also be available.

\item Channel bandwidths should be kept small ($\Delta\nu/\nu \ll 1$)
so as to avoid strong terrestrial transmitters as often occur at low
frequencies and to enable wide-field imaging.

\item Wide-band receiving systems should be employed so that more than
a single frequency can be observed.  Compare the VLA to the CLRO in
Figures~\ref{fig:resolution} and~\ref{fig:sensitivity}.

\item Large, well-filled primary collectors (presumably composed of
numerous individual dipole antennas phased together) should be used so
as to produce a primary beam with high forward gain.

\item Ionospheric calibration should be considered during the design
of the array.  In particular, adequate ionospheric calibration
may require good \emph{aperture}-plane coverage, rather than good
$u$-$v$ plane coverage usually sought in synthesis imaging. Also, the
ability to make simultaneous measurements at widely spaced frequencies
(for phase transfer) and pointing positions (for modelling the ionosphere over
the array) should be available. Rapid (< 10sec) switching may suffice over
true simultaneous measurements, given sufficient sensitivity to track
ionospheric changes on short time scales.
\end{itemize}

\section{Conclusions}\label{sec:conclude}

We have described the 74~MHz system that is installed on the NRAO Very
Large Array and the Pie Town antenna of the Very Long Baseline Array.
All 29 antennas have been outfitted with 74~MHz receivers.  With these
low-frequency receivers and in the VLA's A configuration, the VLA is
the world's highest resolution, highest dynamic range interferometer
operating below~150~MHz.  Working in conjunction with the PT antenna,
the VLA-PT interferometer, with baselines approaching 70~km,
represents the longest baselines ever used for connected element 
synthesis imaging below ~150~ MHz.

The calibration strategy for the 74~MHz VLA incorporates a number of
features not common at higher frequencies.  Chief among these is the
importance of the isoplanatic patch size relative to the primary beam
diameter.  At~74~MHz, the isoplanatic patch size is determined by the
ionosphere and the size of the array, particularly in its larger
configurations, and can be smaller than the primary beam.  We have
developed new \emph{field-based} methods of calibrating 74~MHz VLA
data that are not restricted to assuming that a single phase
correction must apply to the entire field of view.  For certain types
of observations, e.g., a field dominated by a single strong source
that itself is the object of interest, normal, position-independent
self-calibration (such as is used at centimeter-wavelengths) can be
employed to obtain images with reasonable dynamic ranges. 
We have presented snapshot images of 3C
sources of moderate strength as examples of routine, angle-invariant
calibration and imaging, and a sub-sample of these sources with previously
well determined low frequency spectra indicate that the 74 MHz flux scale at
the Very Large Array is stable and reliable to at least 5 percent. The
absolute flux density scale is tied to a model of Cygnus A with a flux
density fixed to the \citet{bgp-tw77} value.

Other aspects of calibration and imaging are not unique to 74~MHz but
assume greater importance relative to higher frequencies.  For
instance, at~1400~MHz one can observe in a pseudo-continuum mode
(i.e., low spectral resolution mode) in order to maintain a large
field of view, to identify RFI, or both.  At~74~MHz, the internal RFI
is sufficiently bad that a pseudo-continuum mode (with somewhat higher
spectral dynamic range than used typically at~1400~MHz) is essential.
Similarly, at~1400~MHz there are typically other sources in the field
of view, and it is possible for there to be strong sources outside the
field of view that must be \textsc{clean}ed in order that their
sidelobes do not reduce the dynamic range in the image.  At~74~MHz,
the presence of many sources in the field of view and of strong
sources outside the field of view is guaranteed.

Although we expect that the 74~MHz system will continue to be a
productive, ``facility-level'' system of the VLA (and EVLA) for many more years,
we also anticipate that the 74~MHz system will be superseded
eventually by LOFAR and the LWA and possibly the
low-frequency end of the Square Kilometer Array (SKA).  Nonetheless,
we believe that it offers many lessons for these future high
resolution telescopes, as
well as the apparatus to explore new science in its own right.

\acknowledgements

We have benefited from discussions with many individuals. A partial
listing includes M.~Bietenholz, K.~Blundell, C.~Brogan, T.~Clarke,
J.~Condon, K.~Dyer, T.~Ensslin, C.~Lacey, W.~Tschager, and K.~Weiler.
The National Radio Astronomy Observatory is a facility of the National
Science Foundation operated under cooperative agreement by Associated
Universities, Inc.  ASC and WML were supported by National Research
Council-NRL Research Associateships.  Basic research in radio
astronomy at the NRL is supported by 6.1 base funding.

\appendix

\section{Snapshot Observations of Bright Sources}\label{sec:example}

The prototype 8-antenna 74~MHz system \citep{kped93} was capable of
imaging only the dozen or so strongest sources in the sky whose flux
densities were hundreds to thousands of Janskys. After the full deployment
of the VLA 74~MHz system, a multi-configuration snapshot survey of
sources from the 3C catalog was started. Observations were obtained in
a succession of three VLA configurations in 1998, A configuration
7-8 March, B configuration 4-5 October, and C configuration 21 November
and 4-5 December. Data in both circular polarizations was obtained
simultaneously in 1 IF each at 74 and 330~MHz. The data were obtained
in spectral line mode with 32 channels at 74~MHz and 64 channels at
330~MHz after online Hanning smoothing. The total available bandwidth
was $\sim$1.5 and $\sim$3~MHz at 74 and 330~MHz, respectively.

All sources were observed numerous times in cycling snapshot fashion to
maximize the hour angle coverage. Our typical scan lengths were 5-10
minutes. Many of the sources were sufficiently small in angular
extent that the A configuration run was sufficient to generate a
good image. More extended sources required B and sometimes C array
data, especially at 330~MHz, because of the higher intrinsic angular
resolution. Table~\ref{tblobs} summarizes the observations and the 
final image beams at 74 and 330~MHz. Notice that in most cases we
convolved the final images with a gaussian, to produce a circular
beam.

The data reduction followed the prescriptions described in the previous
sections of this paper. All of the images were produced from multiple
snapshot observations using an angle-invariant calibration (self-calibration)
strategy, which is sufficient for these strong sources because they dominate
the self-calibration solution and sidelobe confusion can be ignored.
In some cases, work is underway to produce even higher resolution, higher
dynamic range images. Superior images are readily obtained with full synthesis
observations as compared to the snapshot images presented here.

Because of the calibration method used, the locations of the 330 and
74~MHz images are uncertain due to ionospheric wander. We used scaled
subtractions ($330{\rm~MHz-\alpha_{ave}}\times73{\rm~MHz}$) to test for shifts
between the maps and adjusted the 74~MHz images to agree with those at
330~MHz. Note that 330~MHz astrometry could be off by as much as 5\arcsec\
from the true radio reference frame. Alternatively, by using positions
from the NVSS survey one can achieve positional accuracies of $\sim$1\arcsec\
at this frequency.

Figure~\ref{fig:3c} shows 74 and 330~MHz images of a variety of moderately
resolved 3C sources with flux densities on the scales of tens of Janskys or
higher.  Images such as these can now be made routinely with snapshot
observations of tens of minutes or less. In most cases these are the first
sub-arcminute images of these sources below~100~MHz. Table~\ref{tab:list}
reports the peak brightness and flux density of the source, but, for the
resolved sources for which we present images, these values may be lower
limits, as some of the flux may have been resolved out. We discuss the
resolved sources briefly but make no quantitative analyzes, as our
discussion here is intended only to be illustrative.

\begin{description}

\item[{3C~10}]
Figure~\ref{fig:3c}\textit{a} shows the images of Tycho's supernova remnant
at 74 and 330~MHz. We can see in these images a limg-brightened spherical
shell with a diameter of $\sim$8\arcmin, and enhanced emission toward the NE
half of the shell. This structure is similar to that observed at 330 MHz 
and 1.4~GHz by \cite{katz00}, and at 610~MHz by \cite{duim75}.

\item[{3C~33}]
Figure~\ref{fig:3c}\textit{b} presents the 330 and 74~MHz images of this
radio galaxy. The 74~MHz image shows the lobes and fainter emission
between these two structures. The 330~MHz image resolves some of the
details of this source, separating each lobe into 2 components, as well as
showing diffuse emission extending perpendicular to the jet axis in regions
between the lobes and the nucleus. The structures seen in our images are
similar to the ones seen at 1.5~GHz by \cite{leahy91}. A low resolution
160~MHz image of this source \citep{s77} was able to resolve it only
into two components.

\item[{3C~84}]
Figure~\ref{fig:3c}\textit{c} shows the 74~MHz image of the central
regions of the Perseus cluster, which contains a number of strong radio
sources including \objectname[3C]{3C~84}. An enlarged plot of the 330~MHz
Figure~\ref{fig:3c}\textit{d}
image of this radio galaxy shows a strong core, two lobes in the N-S direction
and some extended emission beyond the lobes. These structures are similar to
the ones detected by \cite{pedlar90} at 1.4~GHz, 330~MHz and 151~MHz.
The 74~MHz image does not resolve the nucleus and lobes, but shows some
diffuse emission extending to the NW and SW, corresponding to previous
outbursts (see \cite{fcbkp02} for a more detailed discussion of this
image and the relation between these structures and X-ray holes).

\item[{3C~98}]
Figure~\ref{fig:3c}\textit{e}
The 74~MHz image shows a double
lobe structure with diffuse emission in between. The 330~MHz image
shows a similar structure, although with higher resolution, allowing
the detection of the hotspots and part of the jet. These structures were
previously imaged at higher frequencies (1.4 and 4.8 GHz) by
\cite{Leahy1997} and \cite{Young2005}. A 160~MHz image of this galaxy
was presented by \cite{s77}.

\item[{3C~129}]
Figure~\ref{fig:3c}\textit{f} shows \objectname[3C]{3C~129} at~74~MHz.
A study of the 330~MHz image \citep{lkehp02} have confirmed
the existence of a steep-spectrum ``crosspiece'' at the head of
\objectname[3C]{3C~129}, along the NE-SW direction, perpendicular to
its main tail. This structure was previously detected at 600~MHz
\citep{jag83} and have recently been observed by \cite{lal04} with the
GMRT at 240 and 610~MHz.  We see no indication of this structure at~74~MHz,
but our resolution is considerably lower, so it is probably blended
with the head of \objectname[3C]{3C~129}.

\item[{3C~144}]
Figure~\ref{fig:3c}\textit{g} shows \objectname[3C]{3C~144} (the
\objectname[]{Crab Nebula}, \objectname[]{Tau~A}) at~74~MHz
and 330~MHz. This source can be used as an
amplitude calibrator in more compact configurations (C and D).
The 74~MHz compact source in the center of the nebula is the
\objectname[]{Crab pulsar} (\objectname[PSR]{PSR~B0531$+$21}).
A higher sensitivity 330~MHz image of this source is presented
by  \cite{Frail1995}, while \cite{Bietenholz1997} present
a study of the 74/330 MHz radio spectral index and find evidence
for intrinsic thermal absorption.

\item[{3C~218} (Hydra~A)]
Figure~\ref{fig:3c}\textit{h} present the 330 and 74~MHz images of Hydra A,
where we can see that this radio galaxy has a complex structure, consisting
of several outbursts. \cite{laneetal2004} present a detailed study of this source,
the spectral indices of the different components and their correlation of the
X-ray emission from the cluster of galaxy where it resides.

\item[{3C~219}]
The 74~MHz emission from this radio galaxy (Figure~\ref{fig:3c}\textit{i})
shows two lobes and diffuse emission between them. This emission is resolved
into better detail by the 330~MHz image, where we can see the N and S
hotspots, the jet and some diffuse emission. These structures are similar to
the ones seen at 1.4~GHz by \cite{Clarke1992}.

\item[{3C~274} (Vir~A)]
Figure~\ref{fig:3c}\textit{j} shows \objectname[3C]{3C~274}
(\objectname[]{Vir~A}), one of the sources that can be used as a
primary bandpass and flux density calibrator, in addition to or
instead of \objectname[]{Cyg~A} (particularly if \objectname[]{Cyg~A}
is not above the horizon). Both 330~MHz and 74~MHz images show 
a complex structure, indicative of the interaction between the radio
plasma and the intra cluster medium. A more detailed discussion about
the 330~MHz image of this source is presented by  \cite{oek00}.

\item[{3C~327}]
The 74~MHz image of this galaxy (Figure~\ref{fig:3c}\textit{k})
shows a double structure along the E-W direction. The E component
is elongated, indicating the presence of multiple components. A
low resolution 160~MHz image from \cite{s77} showed only two blobs
along the E-W direction. The structure seen at 74~MHz are confirmed
by the 330~MHz image, where we can clearly see the hotspots and some
diffuse emission, similar to the structure observed at 8.4~GHz
by \cite{Leahy1997}. 

\item[{3C~353}]
The 74~MHz image of this radio galaxy (Figure~\ref{fig:3c}\textit{l})
is broken into 3 components aligned along the E-W direction. Previous
low frequency (160~MHz) images by \cite{s77} were not able to detect
multiple components, detecting only an extended source in the E-W
direction. The 330~MHz image shows the hotspots and diffuse emission in
the lobes, similar to the structure detected by \cite{Baum1988} at 5~GHz.

\item[{3C~390.3}]
In Figure~\ref{fig:3c}\textit{m} we can see that the 74~MHz image of this
radio galaxy is resolved into 2 lobes and diffuse emission associated
with them. This structure is similar to the one detected at 610~MHz 
by radio \cite{jag87}, using WSRT observations. The 330~MHz image shows
the hotspots and diffuse emission in better detail, as well as the nucleus.
The structures seen at this frequency are similar to the ones detected
with 1.4~GHz VLA observations \citep{Leahy1995}.

\item[3C~392]
Figure~\ref{fig:3c}\textit{n} presents the images of this supernova
remnant, which has similar structure at 74 and 330~MHz. The 330~MHz
image shows details similar to the ones seen at 1.4 GHz
\citep{Jones1993,Giacani1997}. Higher resolution, full synthesis images
at both frequencies of this SNR, also known as W44,
are presented in \cite{castellettietal07}.

\item[{3C~405} (Cyg~A)]
Figure~\ref{fig:3c}\textit{o} shows our primary bandpass and flux density
calibrator, \objectname[3C]{3C~405} or \objectname[]{Cyg~A}.  This
image is dynamic range limited. 3C405 is almost unresolved at 74~MHz,
however, higher resolution (VLA $+$ Pie Town) images at both
74 and 330~MHz are presented by \cite{lazio06}. The VLA 330~MHz image
is able to resolve the emission into hotspots and associated diffuse
emission, similar to the structure detected by \cite{pe84} at
1.4~GHz, although without the detection of the nucleus and the jets.

\item[{3C~445}]
Figure~\ref{fig:3c}\textit{p} shows that 74~MHz image of this
radio galaxy is composed of two bright lobes and some faint extended
emission around the nucleus. A lower resolution 160~MHz image of
this galaxy \citep{s77} shows only three blobs. The VLA 330~MHz
image resolves the lobes into hotspots and some associated 
diffuse emission, similar to  that observed at higher frequencies
\citep{Kronberg1986,Leahy1997}.

\item[{3C~452}]
The 74~MHz image of this radio galaxy is presented in
Figure~\ref{fig:3c}\textit{q}, which shows a double lobed structure,
similar to the one detected at 610~MHz by \cite{jag87}, using WSRT.
The 330~MHz image resolves the structure of this galaxy into better
detail, hotspots and diffuse emission along the jet, similar to the
structure detected at 1.4~GHz by \cite{dennett99}.

\item[{3C~461}]
Figure~\ref{fig:3c}\textit{r} shows the 74 and 330~MHz of 
\objectname[]{Cas~A}, another source that can be used as a
primary bandpass and flux density calibrator in more compact
configurations (C and D). The 330 and 74~MHz images look similar,
although of lower resolution compared to the 1.4 GHz images
presented in \citet{Anderson1991}. \cite{Kassim1995} found
evidence of internal thermal absorption using the prototype
8-antenna system, that was subsequently confirmed by \cite{Delaney2004}
using the full 74 MHz VLA + PT Link (Figure~\ref{fig:casapt}).
These observations were obtained with a maximum baseline is ~72 km,
corresponding to an angular resolution of 8.5\arcsec.

\end{description}

\clearpage

\begin{figure}
\caption[]{A picture of a 74~MHz dipole mounted on a VLA
antenna.  In the center of the picture is the subreflector, supported
by the quadrupod legs.  The 74~MHz (crossed) dipoles are in the lower
center of the picture.  The cable that carries the signals from the
dipoles to the receivers drops from the intersection of the dipoles to
the bottom of the antenna's surface.  Also visible just below the
subreflector are the 330~MHz dipoles.}
\label{fig:mount}
\end{figure}

\begin{figure}
\epsscale{0.6}
\caption[]{The block diagram of the 74~MHz receiver on the Pie Town
VLBA antenna.  The receivers on the VLA antennas are similar, with the
main difference being that the VLA receivers are not hetrodyne
receivers.  Rather the 74 and~330~MHz signals are transferred directly
to the intermediate frequency (IF) transmission system.}
\label{fig:receiver}
\end{figure}

\begin{figure}
\rotatebox{-90}{\plotone{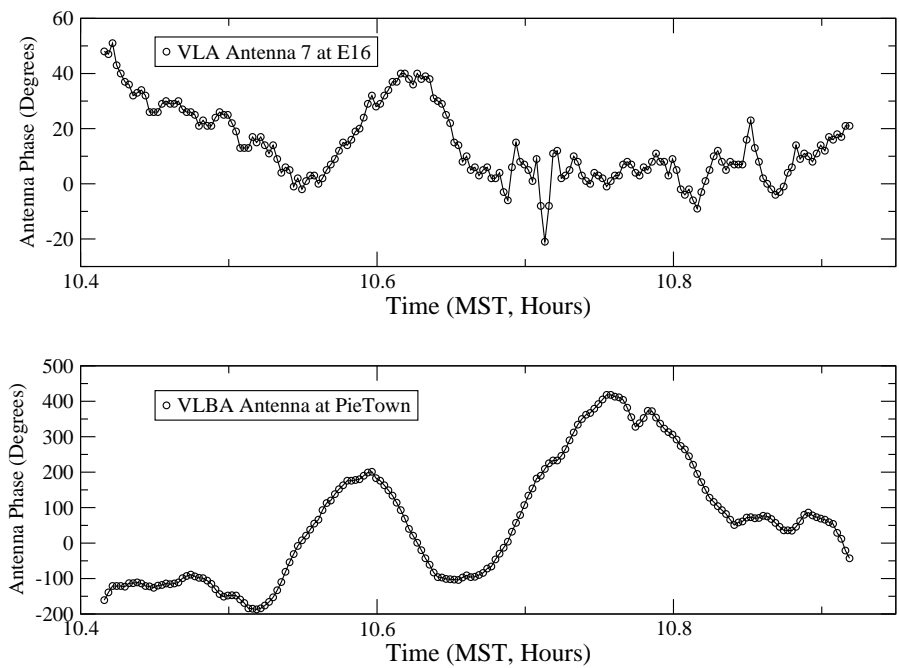}}
\caption[]{First fringes using the PT Link at~74~MHz.  The source
observed was \protect\objectname[3C]{3C~123}.  
\textit{Top} Phase as a function of time for VLA antenna~\#7, located
relatively close to the center of the array.
\textit{Bottom} Phase as a function of time for the PT antenna.  In
both cases, the phases are measured relative to an antenna near the
center of the array.  The scales on the ordinates differ.}
\label{fig:ptlink}
\end{figure}

\begin{figure}
\epsscale{0.85}
\caption[]{The primary beam power pattern at~74~MHz from one of the
antennas.  Other antennas have similar power patterns.  The left panel
shows the left circular polarization, and the right panel shows the
right circular polarization.  The axes are the sines of the offset
angle.  The contours show the decrease, in dB, from the peak of the
pattern (drawn at 3, 6, 10, 15, 20, 25, and 30 dB levels).
The sharp edge on the right hand side of the plots results
from the motion limits on pointing for the VLA antennas.}
\label{fig:beam}
\end{figure}

\begin{figure}
\epsscale{0.7}
\rotatebox{-90}{\plotone{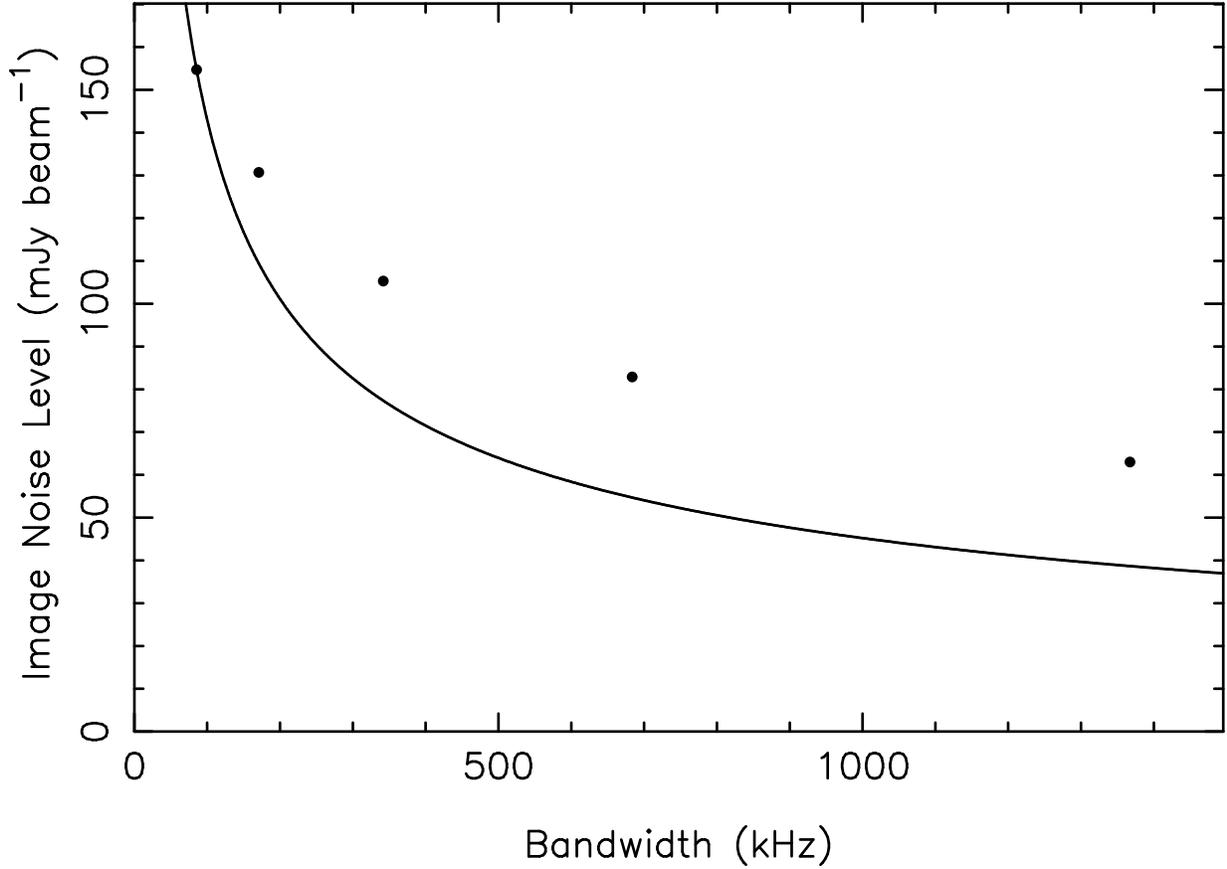}}
\caption[]{The rms image noise level as a function of receiver
bandwidth.  The dots show the measured values from a 90~min.\
integration calibrated and imaged in the fashion described in
\S\S\ref{sec:calibrate} and~\ref{sec:imaging}.  The solid line shows
the behavior expected, $\Delta\nu^{-1/2}$, if the system performance
is limited by thermal noise. This behaviour is typical of observations
obtained in the A and B configurations.}
\label{fig:bandwidth}
\end{figure}

\begin{figure}
\epsscale{0.7}
\rotatebox{-90}{\plotone{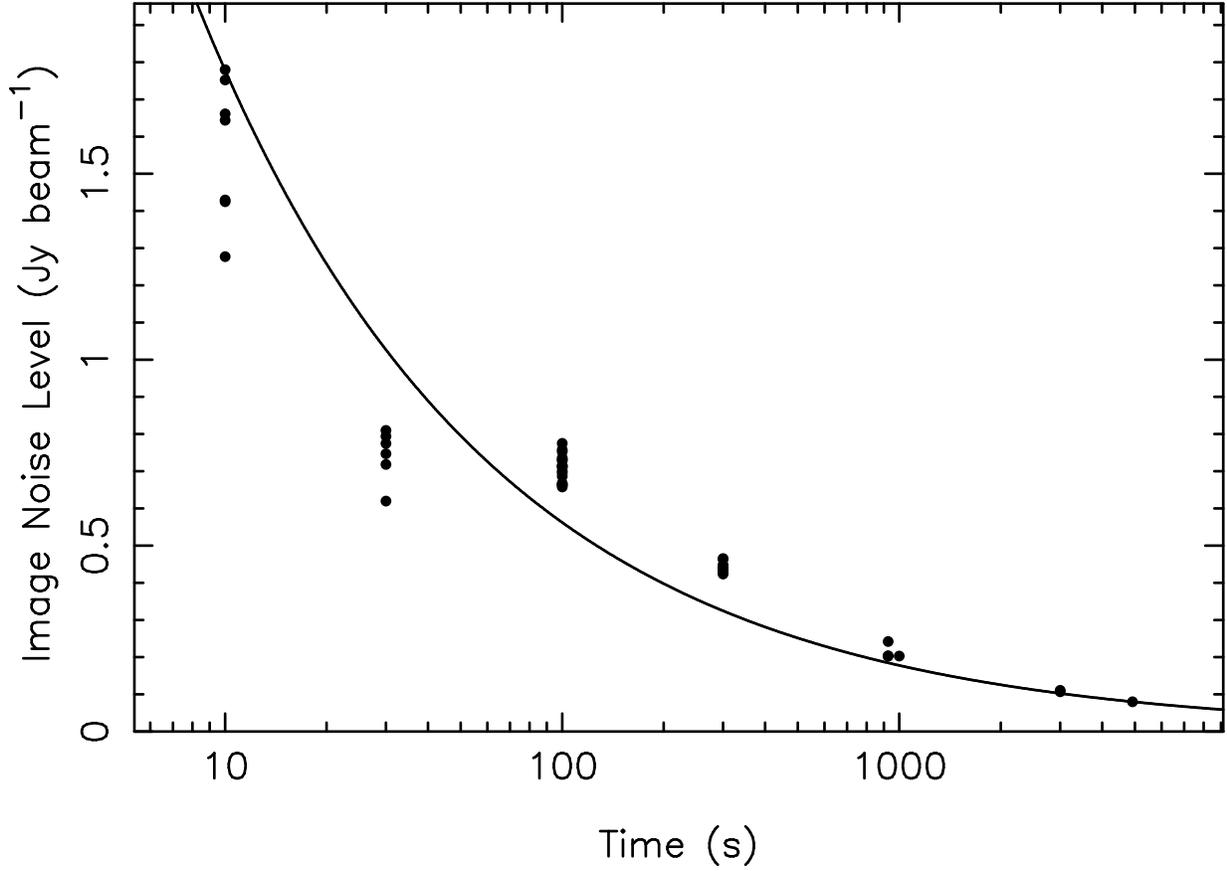}}
\caption[]{The rms image noise level as a function of integration
time.  Multiple points at the same integration time indicate
integration times for which multiple, independent images were
constructed from the approximately 1-hr total integration time.  The
solid line shows the behavior expected, $t^{-1/2}$, if the system
performance is limited by thermal noise. This behaviour is typical of observations
obtained in the A and B configurations.}
\label{fig:time}
\end{figure}

\clearpage

\begin{figure}
\epsscale{0.65}
\caption[]{A demonstration of the excision of 74~MHz \hbox{RFI}.
\textit{Left:} The visibility amplitudes on a single baseline are
displayed in a (linear) gray scale format with frequency on the
abscissa and time on the ordinate.  The time axis is not linear, and
the thick horizontal black stripes indicate times when other sources were
being observed.  The bright vertical stripes represent the 100~kHz
``comb''.  The strength of the comb varies from baseline to baseline
and as a function of time within individual baselines, depending upon
the orientation of the antennas and how strongly they couple.  For
this illustration, we have chosen a particularly severe example; there
were other baselines in the same data set for which the comb was
barely visible.
\textit{Right:} The same data after RFI have been flagged with \task{FLGIT}.
This panel shows a much more homogeneous brightness distribution than the
one to the left, indicating that most of the RFI (strongest interference) was
removed.}
\label{fig:rfi}
\end{figure}

\begin{figure}
\epsscale{0.8}
\plotone{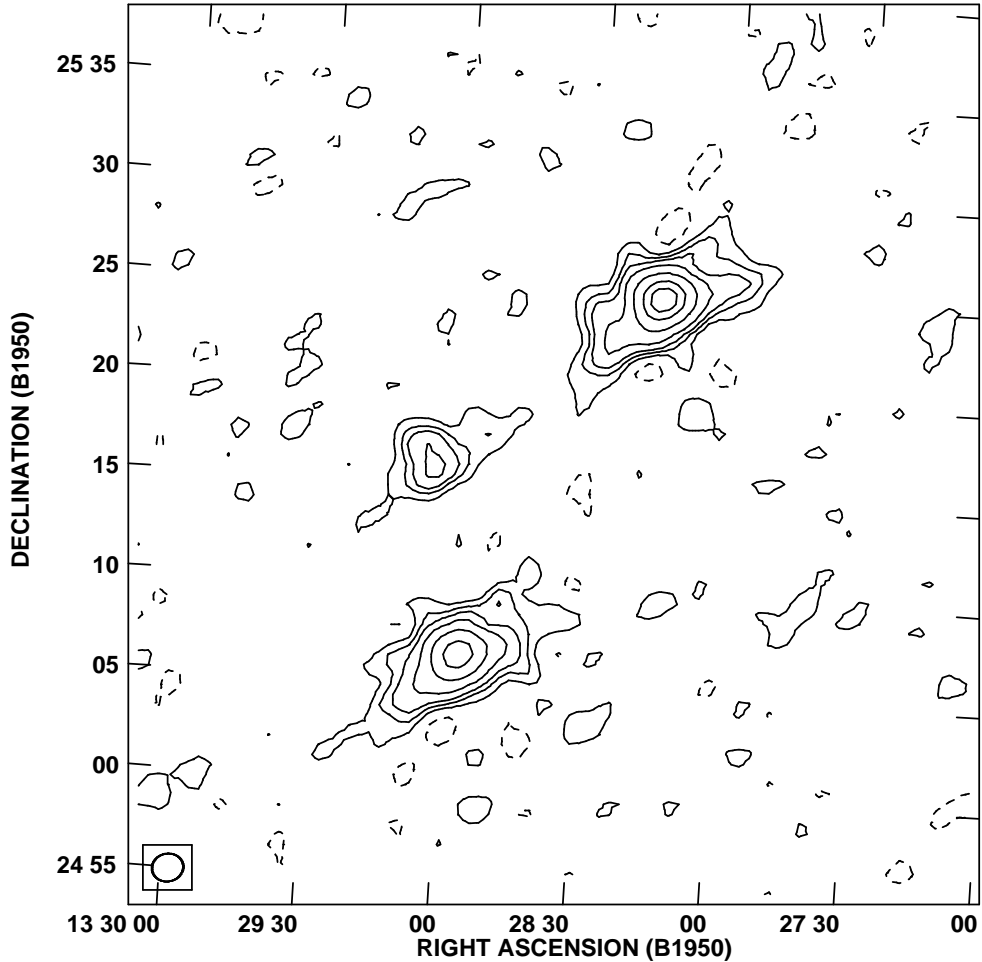}
\caption[]{An example of the distortions introduced by imaging a
region larger than the isoplanatic patch.  This image is a subimage of
a larger image.  The sources shown are roughly 7\arcdeg\ from the
phase center of the larger image; the two brightest sources should be
point-like or nearly so while the third, fainter source is extended,
with a nearly north-south orientation.  The image was produced from
combined B- and C-configuration observations; the beam is 95\arcsec\
$\times$ 83\arcsec\ and is shown in the lower left.  The rms noise
level in the image is approximately 30~\mjybm, and the contour levels
are 30~\mjybm\ $\times -4$, $-2$, 2, 4, 6, 10, 20, 40, 60, 100,
and~150.  The bandwidth and time averaging used in producing this
image are sufficiently small that both contribute a negligible amount
($< 1$ beamwidth) of smearing.}
\label{fig:isoplane}
\end{figure}

\begin{figure}
\epsscale{0.8}
\caption[]{An illustration of the position offsets arising from phase
distortions caused by the largest spatial structures $(\gtrsim
1000$~km) in the ionosphere.  The contours show the 74~MHz image; the
gray scale is an overlay of the NVSS in this region.  The NVSS is at a
sufficiently high frequency that the ionospheric refraction is much
less than a beamwidth.  Note that not all NVSS sources have a 74~MHz
counterpart.}
\label{fig:ionshift}
\end{figure}

\begin{figure}
\epsscale{0.75}
\caption[]{The temporal variation in the refractive shift caused by
the largest spatial scales in the ionosphere.  Shown is the offset, in
both right ascension and declination, of \protect\objectname[]{Cyg~A}
from its known position as a function of time, with a 1~min.\ sampling
interval.  These observations were taken in the B configuration.}
\label{fig:ionwedge}
\end{figure}

\begin{figure}
\epsscale{0.8}
\caption[]{An illustration of the phase distortions caused by
ionospheric mesoscale structure, in this case a traveling ionospheric
disturbance (TID), with the typical scale of order hundreds of
kilometers.  Shown are the phases, measured relative to an antenna
near the center of the array, as a function of time for three antennas
along the west arm of the array.}
\label{fig:iontid}
\end{figure}

\clearpage

\begin{figure}
\epsscale{0.8}
\rotatebox{-90}{\plotone{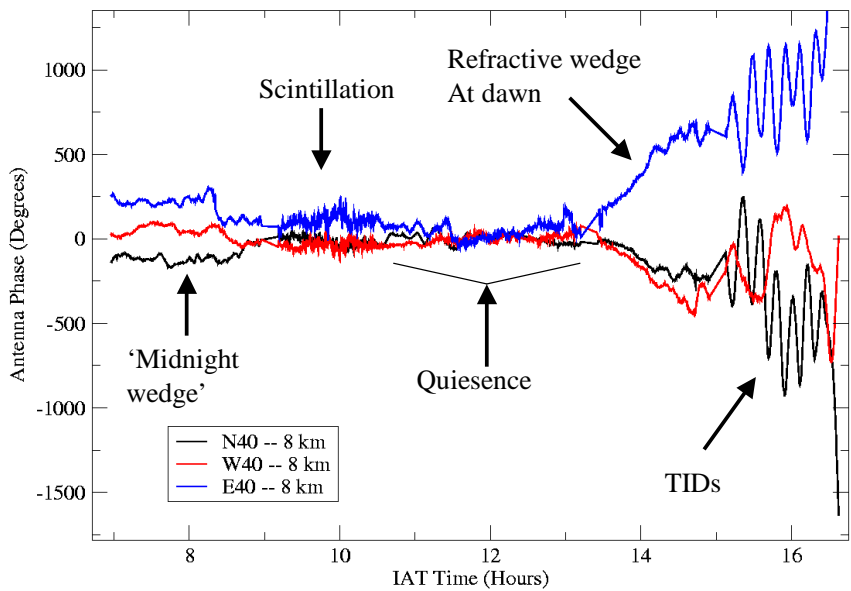}}
\hspace{-8.0cm}
\caption[]{
a) The phase of three antennas relative to a central antenna during an
approximately 8 hr observations of \protect\objectname[]{Vir~A}
illustrating many of the ionospheric phenomena typically observed at
the VLA. All three antennas are located approximately 8 km from the
reference antenna, but represent different azimuths. For the TID the
observed parameters were a period of 750i~s, phase slope of
50~deg~km$^{-1}$, and a time lag of $\sim$50 seconds over 20 km allowing
\cite{p02} to derive a TID wavelength of 750 km and velocity of 200 m~s$^{-1}$;
b) Same as in panel a except for two antennas at different distances
along the same azimuth, indicating that to first order the phase effects of
all the phenomena are proportional to baseline length;
c) The instantaneous amplitude or apparent defocusing of
\protect\objectname[]{Vir~A} over the same time scale as in panel a,
(Hour Angle -4 corresponds to IAT Time$\sim$8~hrs), sunrise
occurred at $+3^{\mathrm{h}}$;
d) The refraction (or apparent postion wander in both RA and Dec) of Virgo
A over the same time scale as panel a;
e,f)  The apparent differential refraction in RA (e) and declination (f) as
measured towards 5 objects located within 6 degrees of
\protect\objectname[]{Vir~A} and of sufficient strength to be detected and
tracked over the course of the observations. The time scale is the same as
in panel c.
}
\label{fig:ionsmall}
\end{figure}

\begin{figure}
\rotatebox{-90}{\plotone{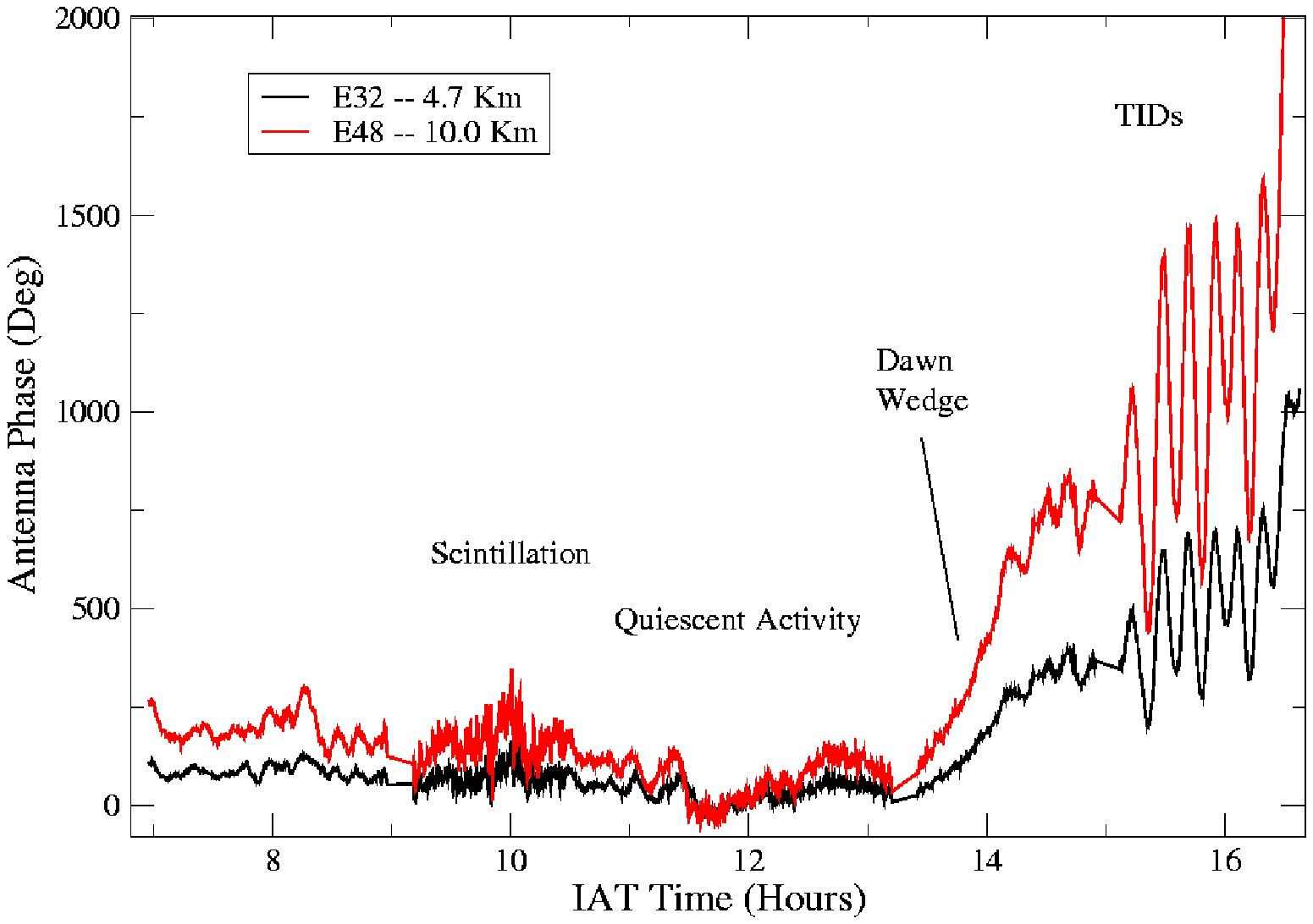}}
\end{figure}

\clearpage

\begin{figure}
\plotone{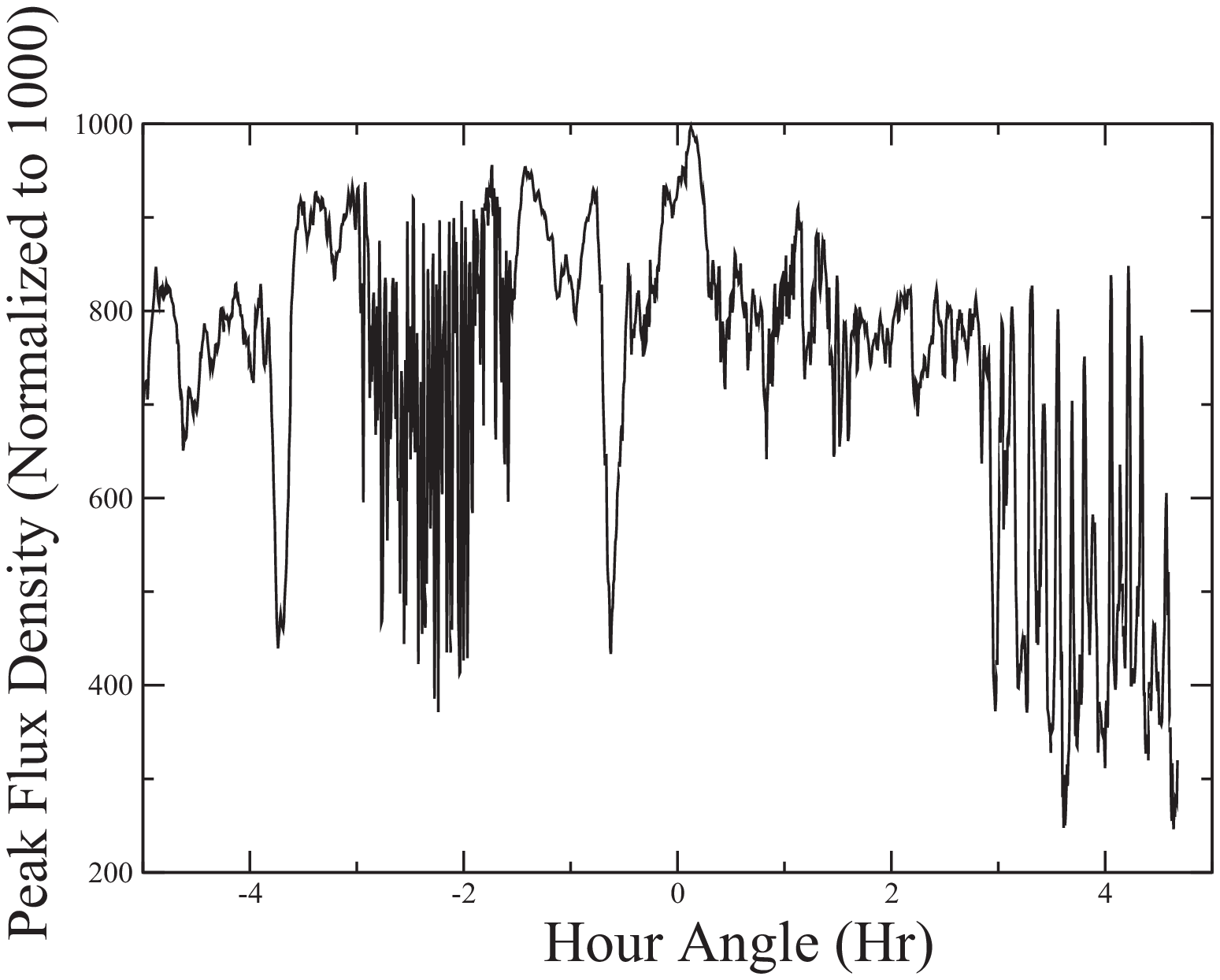}
\end{figure}

\begin{figure}
\rotatebox{-90}{\plotone{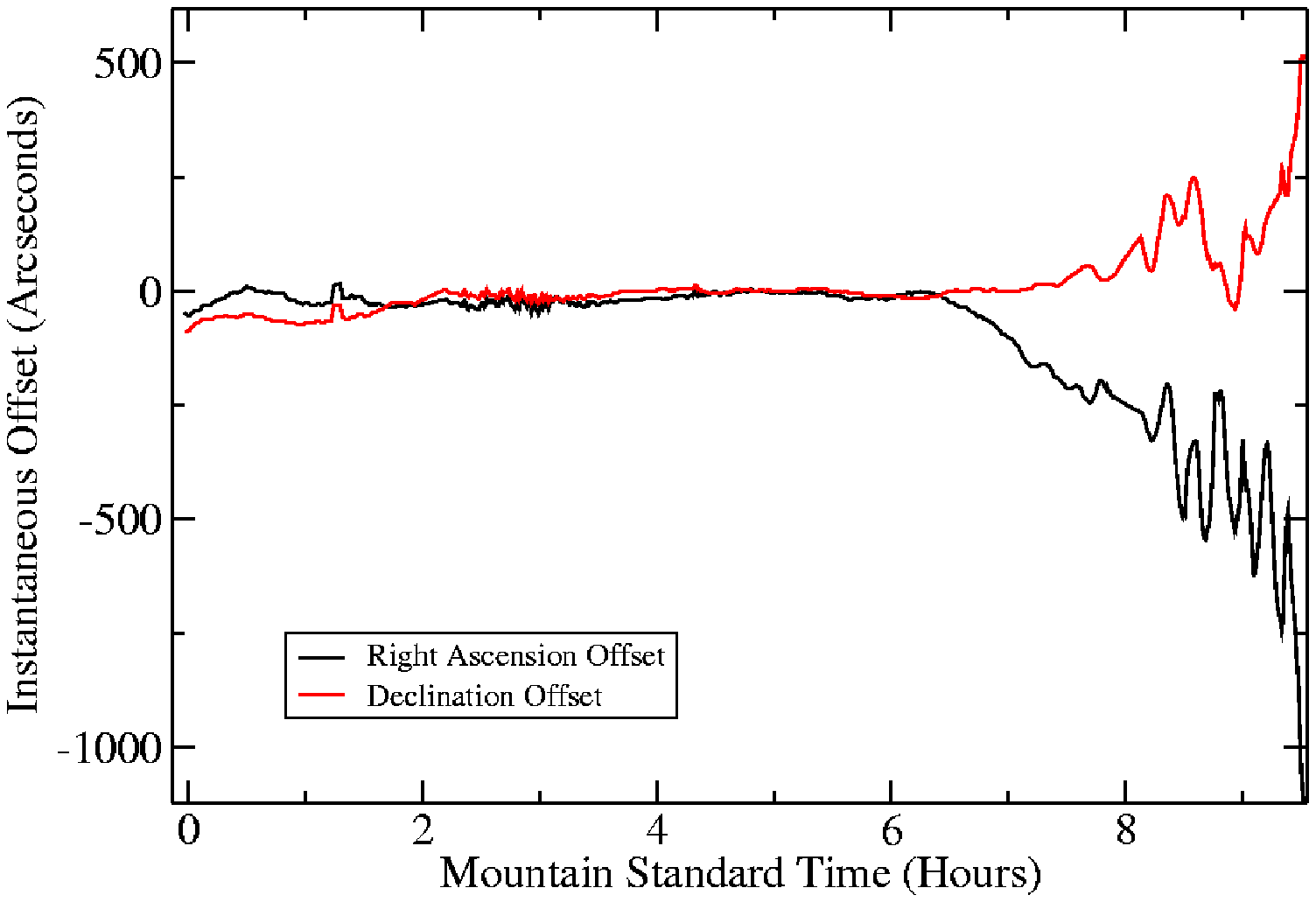}}
\end{figure}

\clearpage

\begin{figure}
\plotone{f14e.eps}
\end{figure}

\begin{figure}
\plotone{f14f.eps}
\end{figure}

\clearpage

\begin{figure}
\epsscale{0.9}
\rotatebox{-90}{\plotone{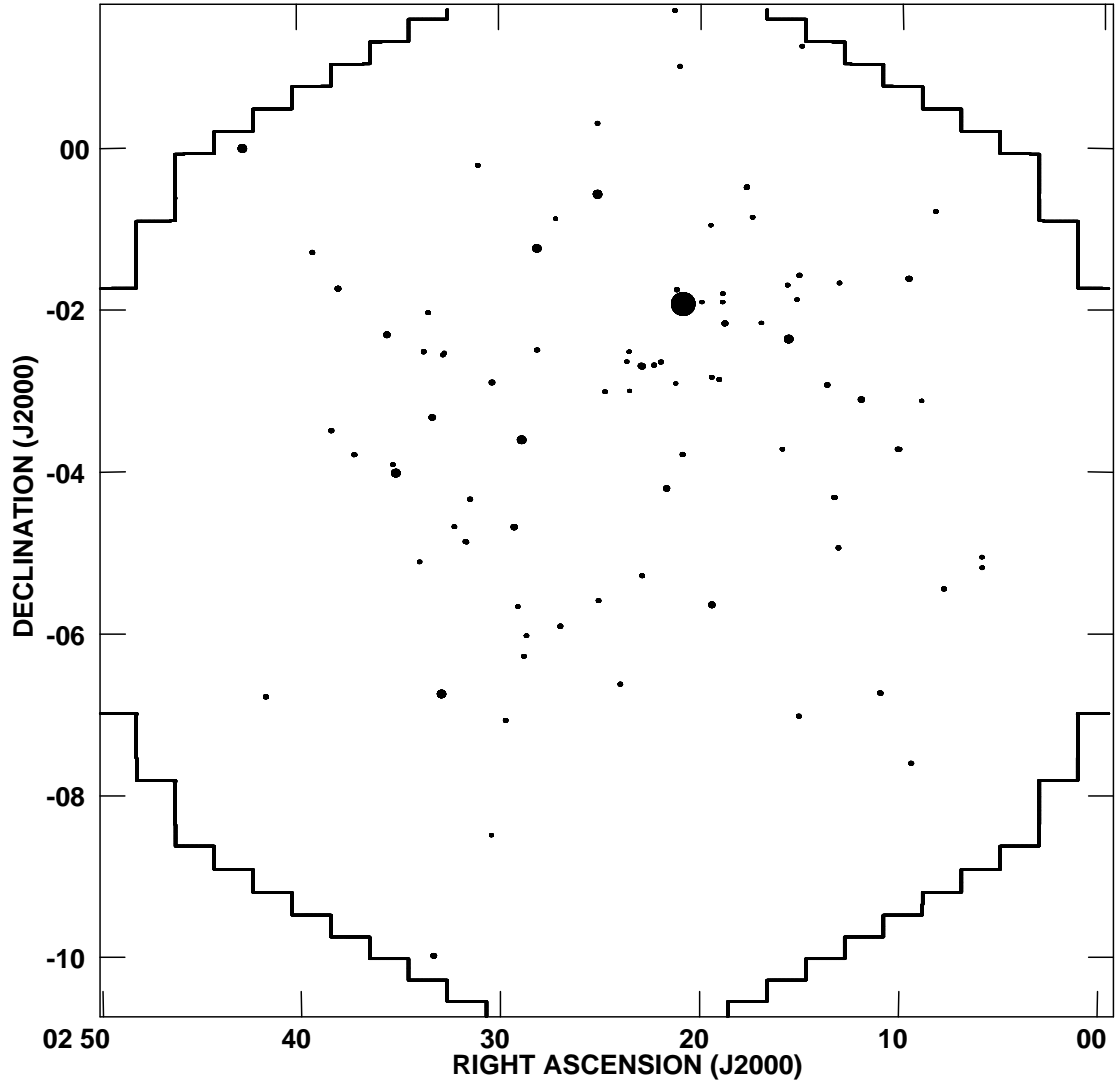}}
\caption[]{An example of the bias introduced by self-calibration.
Symbols show the location of sources and are proportional to their
flux densities.  This field contains the (strong) source
\protect\objectname[3C]{3C~63} in the upper right.  A clear
non-uniform distribution of other sources across the field is evident,
in addition to the decrease in number density expected from the
primary beam attenuation.  The jagged edges in the image indicate the
boundaries of facets (\S\ref{sec:imaging}); regions beyond the edges
of the image have not been imaged.}
\label{fig:SCbias}
\end{figure}

\begin{figure}
\epsscale{0.8}
\rotatebox{-90}{\plotone{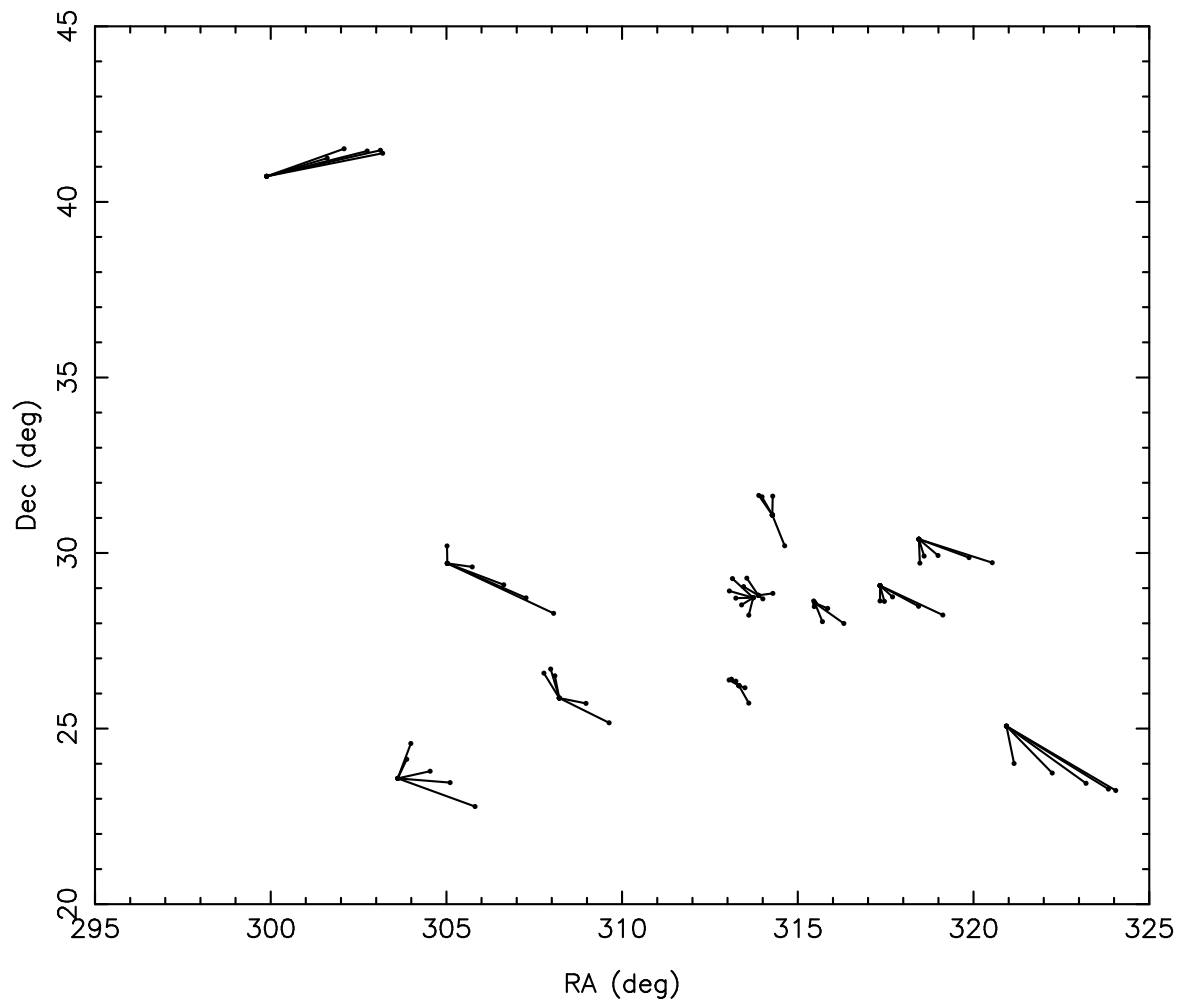}}
\caption[]{Differential, ionospheric-induced source wander within a
field of view.  The expected locations of various moderately strong
sources within a single field of view is shown.  The direction of each
vector indicates direction of the position shift in five 5-min.\
intervals; the length of each vector is 100 times the actual
displacement.  Because the magnitude and direction of the wander is
not the same for all sources, averaging over time results in sources
being smeared out and apparently disappearing from view (viz.\
Figure~\ref{fig:SCbias}).}
\label{fig:wander}
\end{figure}

\clearpage

\begin{figure}
\epsscale{0.9}
 \plottwo{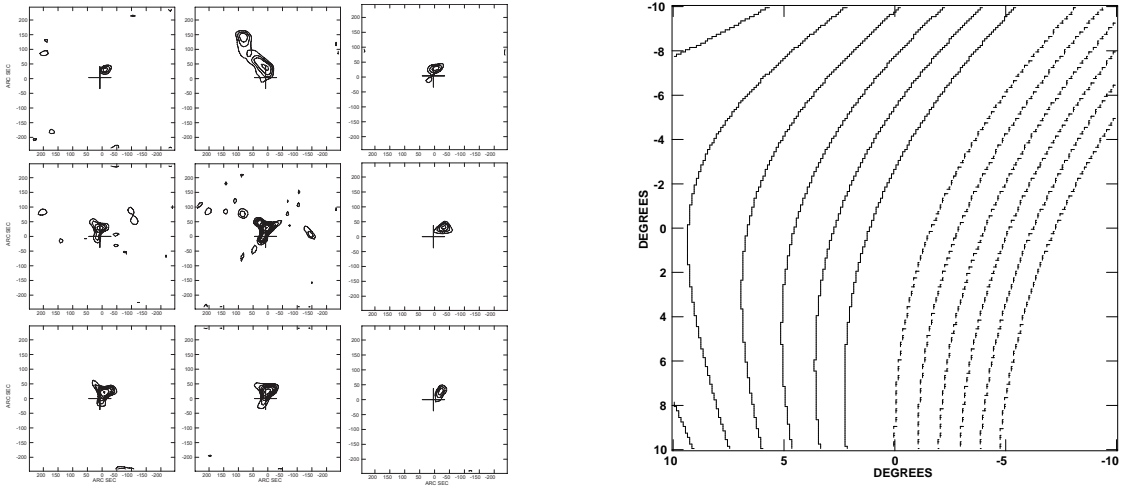}{f17b.eps}
\caption[]{\textit{Left}: A mosaic of NVSS sources assembled from a
1~min.\ snapshot of a full-field image at~74~MHz.  The typical
distance of each source from the phase center is 3\arcdeg.  The
cross in each panel marks the nominal location of the NVSS source.
Offsets from this nominal position are due to ionospheric refraction.
\textit{Right}: The Zernike model for $\phi_{\mathrm{ion,lo}}$ over
the VLA for the same time as the mosaic was constructed.  See
equation~(\ref{eqn:zernike}).  Shown is the phase delay screen above
one VLA antenna; because of the ``small-array'' approximation used,
the phase delay screen over all other antennas is essentially the
same.  At a typical 350~km altitude, the phase delay screen shows
structures on scales of roughly 100~km.  Solid contours represent a
phase advance, relative to a nominal phase, while dashed contours
represent a phase delay.  The most negative contour (lower right)
represents a phase delay of~$-30\arcdeg$, and the most positive
contour (upper left) represents a phase advance of~$+30\arcdeg$.}
\label{fig:zernike}
\end{figure}

\begin{figure}
\epsscale{0.9}
\caption[]{The same field as in Figure~\ref{fig:SCbias}, but
calibrated by treating the ionosphere as a phase-delay screen.  The
distribution of sources across the field is far more uniform than
before, though the primary beam attenuation near the edges of the
image is still apparent.  Abscissa: Right ascension; Ordinate:
Declination.  The jagged edges in the image indicate the boundaries of
facets (\S\ref{sec:imaging}); regions beyond the edges of the image
have not been imaged.  The black circle outside the field of view in
the upper right is the location of an outlier field; this field
contains a strong source outside the field of view that was also
imaged.}
\label{fig:SCunbias}
\end{figure}

\clearpage

\begin{figure}
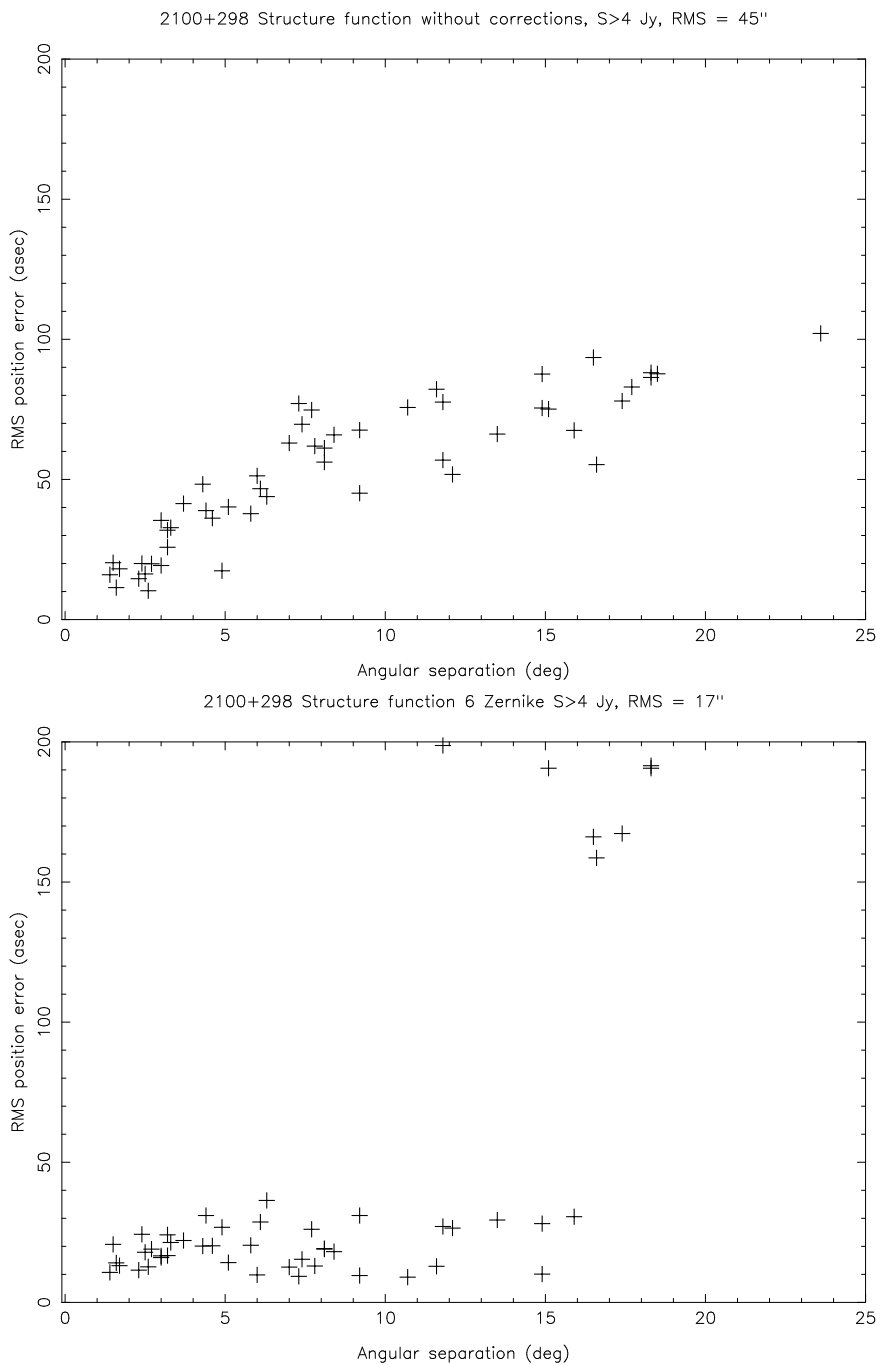

\begin{center}
\epsscale{0.54}
\rotatebox{-90}{\plotone{f19a.eps}}\\
\rotatebox{-90}{\plotone{f19b.eps}}
\end{center}
\vspace{-0.33cm}
\caption[]{
The rms jitter in apparent separations of pairs of sources with
different separations made from a sequence of one minute snapshots
taken over a period of several hours on a day with a moderately
disturbed ionosphere.
\textit{Top} The rms jitter with no correction to the
position difference.
\textit{Bottom} The rms jitter after applying an ionospheric model
described by a 5-term Zernike phase screen determined from
that snapshot.  The large rms values at angular separations greater
than approximately 15\arcdeg\ result from \protect\objectname[]{Cyg~A}
which was not corrected as part of this Zernike modeling.}
\label{fig:sf}
\end{figure}

\clearpage

\figcaption{Various (resolved) 3C sources at~74~MHz (left panel) and 330~MHz
(right panel). For each source, we quote the rms noise level and the beam
diameters in Table~\ref{tblobs}. All beams are circular unless otherwise
noted. The contour  levels are given in terms of the  rms noise level in
the image i($-3$, 3, 6, 12, 24, 48, \ldots times that rms noise level)
and the beam is shown in the lower left.
(\textit{a}) \protect\objectname[3C]{3C~10};
(\textit{b}) \protect\objectname[3C]{3C~33};
(\textit{c}) The central strong sources of the \protect\objectname[]{Perseus
cluster} at~74~MHz.  The source in the lower left is
\protect\objectname[3C]{3C~84}, the compact source near the center is
\protect\objectname[3C]{3C~83.1A}, and the extended radio galaxy in
the upper right is \protect\objectname[NGC]{NGC~1265};
(\textit{d}) \protect\objectname[3C]{3C~84};
(\textit{e}) \protect\objectname[3C]{3C~98};
(\textit{f}) \protect\objectname[3C]{3C~129} (western source) and \protect\objectname[3C]{3C~129.1} (eastern source);
(\textit{g}) \protect\objectname[3C]{3C~144};
(\textit{h}) \protect\objectname[3C]{3C~218};
(\textit{i}) \protect\objectname[3C]{3C~219};
(\textit{j}) \protect\objectname[3C]{3C~274} or \protect\objectname[]{Vir~A};
(\textit{k}) \protect\objectname[3C]{3C~327};
(\textit{l}) \protect\objectname[3C]{3C~353};
(\textit{m}) \protect\objectname[3C]{3C~390.3};
(\textit{n}) \protect\objectname[3C]{3C~392};
(\textit{o}) \protect\objectname[3C]{3C~405} or \protect\objectname[]{Cyg~A};
(\textit{p}) \protect\objectname[3C]{3C~445};
(\textit{q}) \protect\objectname[3C]{3C~452};
(\textit{r}) \protect\objectname[3C]{3C~461}.
\label{fig:3c}}

%
%
%
%
%
%
%
%
%
%
%
%
%
%
%
%
%
%

\begin{figure}
\caption[]{VLA + PT Link full synthesis 74~MHz image of Cas A, with a resolution of
$\sim$8\arcsec. The bottom bar shows the color transfer function in Jy.
Image courtesy T. Delaney \cite{Delaney2004}}
\label{fig:casapt}
\end{figure}

\clearpage

\begin{deluxetable}{lr}
\tablecaption{Performance Characteristics of the VLA 74~MHz
	System\label{tab:perform}}
\tablewidth{0pc}
\tablehead{}
\startdata
Center Frequency                  & 73.8~MHz \\
Bandwidth                         & 1.6~MHz  \\
Primary Beamwidth (FWHM)          & 11\fdg7  \\
System Temperature (minimum)      & 1500~K   \\
Aperture Efficiency               & $\approx 15$\% \\
Point Source Sensitivity (A or B configs., 8~hr) & 25$-$50~mJy\\
Resolution (A configuration)      & 25\arcsec \\
\phantom{Resolution} (VLA $+$ PT) & 12\arcsec \\
\enddata
\tablecomments{See also Figures~\ref{fig:beam}, \ref{fig:bandwidth},
and~\ref{fig:time}.}
\end{deluxetable}

\begin{deluxetable}{lcccl}
\tablewidth{0pc}
\tablecaption{Comparison of VLA and \citeauthor{kwp-tn81} Flux Density
	Estimates\label{tab:fluxscale}}
\tabletypesize{\footnotesize}
\tablehead{
 \colhead{Name} & \colhead{\citeauthor{kwp-tn81}} & \colhead{VLA} &
	\colhead{Ratio} & \colhead{Remarks} \\
                & \colhead{(Jy)}            & \colhead{(Jy)}
}
\startdata

\objectname[3C]{3C~48}    & 69.1  & 67.6 $\pm$ 0.4   & 0.978 &
	\citeauthor{kwp-tn81} spectrum fit below observations \\
			   	    		       	        
\objectname[3C]{3C~98}    & 98.9  & 98.0 $\pm$ 1.1   & 0.991 & \\
			   	    		       	        
\objectname[3C]{3C~123}   & 387.6  &  414.7 $\pm$ 1.3 & 1.070 & 
	Both \citeauthor{kwp-tn81} spectrum and VLA data look good \\
			   	    		       	        
\objectname[3C]{3C~147}   &  67.1 &  55.3 $\pm$ 0.8  & 0.825 &
	\citeauthor{kwp-tn81} estimate could be high, only one datum \\
                          &        &                  &       &
	below~178~MHz and spectrum fit well below it \\
			   	    		       	        
\objectname[3C]{3C~219}   &  96.6 &  94.8 $\pm$ 0.5  & 0.981 & 
	\citeauthor{kwp-tn81} spectrum looks good \\

\\

\objectname[3C]{3C~274}   & 2281.3 & 2084.6 $\pm$ 1.3    & 0.914 & 
	Both \citeauthor{kwp-tn81} spectrum and VLA data look good \\

\objectname[3C]{3C~327}   &  93.1 & 118.4 $\pm$ 0.8  & 1.272 &
	\citeauthor{kwp-tn81} spectrum could be low, \\
                          &        &                  &       &    
	falls below~38~MHz datum but fits 80~MHz datum \\
			   	    		       	        
\objectname[3C]{3C~353}   & 437.0  & 443.8 $\pm$ 0.8  & 1.016 &
	\citeauthor{kwp-tn81} spectrum is unreliable below~160~MHz \\
			   	    		       	        
\objectname[3C]{3C~390.3} & 107.2  &  97.1 $\pm$ 0.4  & 0.906 &
	\citeauthor{kwp-tn81} spectrum may be high, \\
                          &        &                  &       &
	misses 38~MHz datum \\
			   	    		       	        
\objectname[3C]{3C~445}   &  62.6 & 49.2 $\pm$ 0.3 & 0.785 & 
	VLA flux density unestimated?; B-configuration data only \\

\\

\objectname[3C]{3C~452}   &  142.5 & 142.6 $\pm$ 0.2  & 1.001 & \\

\enddata
\tablecomments{The uncertainties given for the VLA measurements are
merely the formal standard deviations from fits to the images.  Little
significance should be attached to them.}

\end{deluxetable}

\begin{deluxetable}{ccccc}
\tablecaption{Zernike Polynomials $Z^l_n(\rho, \phi)$\label{tab:zernike}}
\tablewidth{0pc}
\tablehead{
		      & \multicolumn{4}{c}{$l$} \\
	\colhead{$n$} & 0 & 1 & 2 & 3}
\startdata

1 & \nodata & $\rho e^{i\phi}$ & \nodata & \nodata \\
2 & $2\rho^2 -1$ & \nodata & $\rho^2 e^{2i\phi}$ & \nodata \\
3 & \nodata & $(3\rho^3 - 2\rho) e^{i\phi}$ & \nodata & $\rho^3 e^{3i\phi}$ \\
\enddata
\tablecomments{The polynomials are expressed in terms of a radial
distance from the phase center~$\rho$ and a position angle~$\phi$.
The $Z^0_0$ polynomial is not used because it represents a total phase
contribution (``piston'') to which the interferometer is insensitive.}
\end{deluxetable}

\begin{deluxetable}{llllll}
\tablecaption{Evolving Techniques for Ionospheric Calibration\label{tab:ioncalib}}
\tablewidth{0pc}
\tablehead{\colhead{Method}&\colhead{Parameterization}&
\colhead{Imaging}&\colhead{FoV}&\colhead{Baseline}&\colhead{Reference}\\
\colhead{}&\colhead{}&\colhead{Capability}&\colhead{}&\colhead{}}
\startdata
Simple &None – re-register &No restrictions&Full field &$\leq$5 km&\cite{e84}\\
 geometric shift&position of known&&($\lambda$/D$_{station}$)&&\\
& sources&&&&\\
&&&&&\\
Classical& $\phi_i$(t) - one term& Bright, isolated& $\leq$15\arcmin&$\leq$400 km&\cite{kped93}\\
Self Calibration&per i station&sources (e.g.& &&\cite{Gizani2005}\\
&& 3C objects)& &&\cite{lazio06}\\
&&&&&\\
Field-based &$\phi$(t,$\alpha$,$\delta$) - single & No restrictions& Full field&$\leq$12 km&\cite{cotton04}\\
 calibration&term for entire&(used for VLSS)&($\lambda$/D$_{station}$)&&\\
&array&&&&\\
&&&&&\\
Joint &$\phi$(t,$\alpha$,$\delta$) -  one& No restrictions& Full field&$\leq$400 km& TBD\\
Multi-source&term per&&($\lambda$/D$_{station}$)&&\\
Self-calibration&i stations&&&&\\
\enddata
\end{deluxetable}

\begin{deluxetable}{ccc}
\tablecaption{74~MHz VLA Classical Confusion
	Limits\label{tab:confuse}}
\tablewidth{0pc}
\tablehead{
	\colhead{Configuration} & \colhead{Resolution}  &
	\colhead{Confusion Level} \\
	                        & \colhead{(\arcsec)} &
	\colhead{(mJy)}}
\startdata
A &  25 &    7 \\
B &  77 &   40 \\
C & 240 &  250 \\
D & 744 & 1500 \\
\enddata

\end{deluxetable}

\begin{deluxetable}{ccc}
\tablecaption{74~MHz VLA Polyhedral Imaging\label{tab:facet}}
\tablewidth{0pc}
\tablehead{
	\colhead{Configuration} & \colhead{$\theta_{\mathrm{facet}}$}  &
	\colhead{$N_{\mathrm{facet}}$} \\
	                        & \colhead{(\arcmin)}}
\startdata
A & 13 & 720 \\
B & 22 & 250 \\
C & 39 & \phantom{1}80 \\
D & 69 & \phantom{1}25 \\
\enddata
\end{deluxetable}

\begin{deluxetable}{llrrrrrrrrr}
\tabletypesize{\scriptsize}
\tablecaption{74~MHz and 330~MHz Observations of Sources\label{tblobs}}
\tablewidth{0pt}
\tablehead{
\colhead{Source}&\colhead{Alternative}&\multicolumn{3}{c}{Integration Time}&\colhead{} &\multicolumn{2}{c}{74~MHz}&\colhead{} &\multicolumn{2}{c}{330~MHz}\\
\cline{3-5} \cline{7-8} \cline{10-11}\\
\colhead{} & \colhead{Name} & \colhead{A-conf.}& \colhead{B-conf.}& \colhead{C-conf.}&\colhead{} &\colhead{beam}&\colhead{rms}&\colhead{} &\colhead{beam}&\colhead{rms}\\
\colhead{}&\colhead{}&\colhead{(min)}&\colhead{(min)}&\colhead{(min)}&\colhead{} &\colhead{(\arcsec)}&\colhead{(Jy)}&\colhead{} &\colhead{(\arcsec)}&\colhead{(mJy)}}
\startdata
3C010   &Tycho SNR         &71 &80 &35  && 80&0.36 &&7.9$\times$6.7&2.0\\
3C033   &                  &41 &62 &29  && 25&0.14 &&7.0&7.4\\
Perseus Cluster&           &51 &73 &111 && 94&0.16 &&20.0&6.7\\
3C084   &NGC1275           &51 &73 &111 && 25&0.08 &&6.0&2.6\\
3C098   &                  &44 &53 &36  && 25&0.13 &&8.0&3.9\\
3C129   &                  &13 &93 &87  && 83$\times$75&0.10&&65.0&6.5 \\
3C144   &M1, Crab SNR      &92 &14 &19  && 25&0.12 &&18.0&47.4\\
3C218   &Hydra A           &16 &16 &18  && 30&0.17 &&10.0&26.0\\
3C219   &                  &82 &82 &58  && 25&0.06 &&6.0&1.9\\
3C274   &M87, Virgo A      &71 &71 &49  && 25&0.08 &&23.0&13.8\\
3C327   &                  &52 &51 &21  && 25&0.38 &&6.0&4.3\\
3C353   &                  &51 &51 &27  && 25&0.16 &&7.5&7.3\\
3C390.3 &                  &72 &81 &27  && 29$\times$25&0.05&&7.5&2.4 \\
3C392   &W44, SNR G34.7-0.4&51 &51 &27  && 300&1.10 &&25.0$\times$22&15.6 \\
3C405   &Cyg A             &9  &9  &10  && 31$\times$26&4.63 &&5.0&53.0\\
3C445   &                  &31 &42 &\nodata&& 30&0.03 &&10.4$\times$9.0&3.0\\
3C452   &                  &76 &82 &33  && 25&0.12 &&7.0&2.8\\
3C461   &Cas A             &7  &9  &7   && 30&17.0 &&18.0&227.0\\
\enddata
\end{deluxetable}

\begin{deluxetable}{lcc}
\tablecaption{Peak Intensities and Flux Densities of 3C Sources
	at~74~MHz\label{tab:list}}
\tablewidth{0pc}
\tablehead{
 \colhead{Name} & \colhead{$I$} & \colhead{$S$} \\
                & \colhead{(Jy~beam${}^{-1}$)} & \colhead{(Jy)}}
\startdata

\objectname[3C]{3C~10}  &  13.0  &  252.1 $\pm$ 1.94   \\
\objectname[3C]{3C~33}  &  55.3  &  105.3 $\pm$ 0.41   \\
\objectname[3C]{3C~48}  &  66.9  &   67.6 $\pm$ 0.44	   \\
\objectname[3C]{3C~84}  &  82.2  &  171.3 $\pm$ 0.52   \\
\objectname[3C]{3C~98}  &  35.7  &   98.0 $\pm$ 1.12	   \\
							     
\\							     
							     
\objectname[3C]{3C~123} &  387.8 &  414.7 $\pm$ 1.31 \\
\objectname[3C]{3C~129} &  27.8  &   70.4 $\pm$ 0.48   \\
\objectname[3C]{3C~144} &  305.6 & 1811.3 $\pm$ 3.07 \\
\objectname[3C]{3C~147} &  57.2  &   55.3 $\pm$ 0.82   \\
\objectname[3C]{3C~161} &  85.6  &   87.5 $\pm$ 0.39   \\
							     
\\							     
							     
\objectname[3C]{3C~196} &  133.1 &  129.8 $\pm$ 0.95 \\
\objectname[3C]{3C~218} &  270.2 &  644.2 $\pm$ 1.83 \\
\objectname[3C]{3C~219} &  38.9  &   94.8 $\pm$ 0.50   \\
\objectname[3C]{3C~273} &  142.2 &  140.6 $\pm$ 1.42   \\
\objectname[3C]{3C~274} &  567.2 & 2084.6 $\pm$ 1.29 \\
							     
\\							     
							     
\objectname[3C]{3C~286} &  27.6  &   27.2 $\pm$ 0.76   \\
\objectname[3C]{3C~295} &  111.6 &  107.9 $\pm$ 1.62 \\
\objectname[3C]{3C~298} &  95.3  &   92.4 $\pm$ 1.22   \\  
\objectname[3C]{3C~327} &  52.8  &  118.4 $\pm$ 0.77  \\
\objectname[3C]{3C~353} &  152.2 &  443.8 $\pm$ 0.76 \\
							   
\\							     
							     
\objectname[3C]{3C~380} &  124.4 &  143.7 $\pm$ 2.57 \\
\objectname[3C]{3C~390.3} & 45.4 &   97.1 $\pm$ 0.42 \\
\objectname[3C]{3C~392} &   9.2  &  715.7 $\pm$ 2.88   \\
\objectname[3C]{3C~405} & 9308.3 & 17205.0 $\pm$ 1.44 \\
\objectname[3C]{3C~445} &  9.2   &  49.2 $\pm$ 0.32	 \\
							   
\\							     
							     
\objectname[3C]{3C~452} &  38.1   &  142.6 $\pm$ 0.19   \\
\objectname[3C]{3C~461} &  2362.2 &  17693.9 $\pm$ 12.12 \\
\objectname[3C]{3C~468.1} & 42.0  &  40.7 $\pm$ 0.66 \\
							  
\enddata
\end{deluxetable}

\end{document}